\documentclass{emulateapj}
\def\ergs{ergs~s$^{-1}$}
\def\ergcms{ergs$^{-1}$~cm$^{-2}$~s$^{-1}$}

\def\hst{{\it HST}}

\def\cha{{\it Chandra}}
\def\xmmn{{\it XMM-Newton}}
\def\xmm{{\it XMM}}

\def\asca{{\it ASCA}}

\begin{document}

\title{Spectral States and Evolution of Ultraluminous X-ray Sources}

\author{Hua Feng\altaffilmark{1,2} and Philip Kaaret\altaffilmark{2}}

\altaffiltext{1}{Department of Engineering Physics and Center for Astrophysics, Tsinghua University, Beijing 100084, China; hfeng@tsinghua.edu.cn}
\altaffiltext{2}{Department of Physics and Astronomy, University of Iowa, Van Allen Hall, Iowa City, IA 52242, USA}

\shortauthors{Feng and Kaaret}
\shorttitle{Spectral Evolution of ULXs}

\begin{abstract}
We examined spectral evolution in ultraluminous X-ray sources (ULXs) with apparent luminosities of about $10^{40}$~\ergs. Based on new results in this paper and those reported in the literature, two common spectral behaviors were found.  Some ULXs in starburst galaxies have varying luminosity ($L$) but remain in the hard state with power-law spectra and a constant, hard photon index ($\Gamma$). Other ULXs, such as NGC 5204 X-1, show a correlation between $L$ and $\Gamma$. We interpret this $L - \Gamma$ correlated phase as an intermediate state with hybrid properties from the thermal dominant and steep power-law states. When the spectra of NGC 5204 X-1 are fitted with a multicolor disk blackbody plus power-law model, the X-ray luminosity increases with the effective temperature of the accretion disk in a manner similar to that found in stellar-mass black hole X-ray binaries, suggesting that the emission arises from an accretion disk. The luminosity, disk size, and temperature suggest that NGC 5204 X-1 harbors a compact object more massive than stellar-mass black holes.  In contrast, the disk model in IC 342 X-1 is ruled out because the luminosity decreases as the temperature increases; sources with such behaviors may represent a class of objects with super-Eddington accretion. Also, we report a peculiar soft spectral feature from IC 342 X-2 and variability on a time scale of 20~ks from Holmberg II X-1. More observations are needed to test these results.
\end{abstract}


\section{Introduction}

Ultraluminous X-ray sources (ULXs) are point-like, nonnuclear X-ray objects with apparent luminosities (inferred assuming isotropic emission) above the Eddington limit of a stellar-mass black hole, typically $3\times10^{39}$~\ergs\ for a 20 $M_\sun$ black hole. Those with rapid variability must be accreting compact objects \citep[e.g.,][]{str03,str07}. Their high luminosities suggest they might be powered by accretion onto black holes with masses in the range of $10^2$--$10^4$ $M_\sun$, i.e.\ the missing population of intermediate-mass black holes \citep[IMBHs;][]{col99,mak00,kaa01}. Alternatively, they could be a special class of stellar-mass black hole binaries, different from most ones in our Galaxy, with beamed and/or super-Eddington emission \citep[][]{fab01,kin01,wat01,beg02,beg06,pou07}. Although their physical nature is still unclear, ULXs have shown diverse phenomena and may be a heterogeneous population \citep{swa04,fen05,fen06b}.

An important argument that supports the IMBH scenario is the discovery of soft excesses in the energy spectra of ULXs, which can be interpreted as emission from an accretion disk \citep{kaa03}. At a fixed fraction of Eddington luminosity, the disk luminosity is proportional to the black hole mass, while its inner temperature is scaled with the mass to a power of $-\case{1}{4}$ \citep{mak00}. Disks in ULXs have inner temperatures in the range of 0.1--0.4 keV \citep{mil04a,fen05,sto06}, cooler but more luminous than in stellar-mass black holes, suggesting that the emission arises from accretion onto IMBHs. However, the cool accretion disk model has aroused broad dispute. \citet{gon06} argued that the soft excess could be a soft deficit depending on how the power-law continuum is modeled. \citet{ber08} found similar soft excesses exist in both ULXs and less luminous objects, implying that they are not a signature of IMBHs. \citet{kin04} and \citet{pou07} suggested that the cool thermal emission is associated with massive outflow in super-Eddington accretion rather than from the inner disk. Due to the quality of the data, some spectra could be equally well fitted by other models that do not need the presence of IMBHs \citep{sto06,vie06,miz07}. Study of multiple spectra obtained at different luminosities from individual ULXs enables one to test if the spectral parameters  evolve as expected in the models.  This provides a means to resolve the ambiguities in fitting individual spectra and determine which models correctly describe the physical behavior of ULXs.

Due to the nature of blackbody emission, an optically thick accretion disk should follow the $L \propto T^4$ relation, where $L$ is luminosity and $T$ is the temperature. This has been widely verified in black hole binaries \citep{gie04}. \citet{fen07a} found that the thermal temperature of NGC 1313 X-2 failed to follow the $L \propto T^4$ relation while the luminosity changed by a factor of 10, providing solid evidence against the disk model. However, it is interesting that the spectral evolution of NGC 1313 X-2 shows a pattern totally different from what was observed in NGC 1313 X-1 \citep{fen06a}. The cool disk model, albeit not favored statistically and proved wrong in one case, still could be correct in some other ULXs. Finding spectral evolution with a pattern of $L \propto T^4$ would be evidence in favor of the cool disk model.

Spectral evolution could also help us understand the spectral states of ULXs. Black hole binaries may show similar spectral features in different emission states. Thus, without timing or multiwavelength information, the spectral state often cannot be unambiguously determined with a snapshot observation. Multiple observations showing an evolution pattern would be more convincing in determining the spectral state. 

We investigated spectral evolution in eight ULXs, NGC 5204 X-1, Holmberg II X-1, IC 342 X-1 and X-2, and Antennae X-11, X-16, X-42, and X-44, using archival and proprietary \xmmn\ observations, as well as \cha\ data if necessary. These sources have peak luminosities over $10^{40}$~\ergs\ and strong long-term variability, thus are good candidates for spectral evolution search. Some ULXs like NGC 5408 X-1, though of great interest, have shown roughly constant spectral parameters in the literature and are not included in the paper. The details of observations and data analysis are described in \S~\ref{sec:obs}. In \S~\ref{sec:diss}, we discuss our results together with results on the spectral evolution of other ULXs with reported in the literature. A brief conclusion of the paper is in \S~\ref{sec:con}.

\section{Observations and Data Analysis}
\label{sec:obs}

We used SAS 7.1.0 with calibration files current as of 2008 July for \xmm\ data reduction. New event files were created from Observation Data Format (ODF) files for each exposure. Multiple exposures made with the same Charge Coupled Device (CCD) in the same observation were combined into a single spectrum (data from different CCDs are not combined). Time intervals were excluded if flares occurred in the 10--15 keV lightcurve of the whole CCD. Energy spectra were extracted from events with FLAG equal 0 and PATTERN no more than 4 for PN or 12 for MOS. For each observation, we used the same set of source and background regions in sky coordinates to extract spectra from different CCDs. The source extraction region was defined as a circle centered on the source with a radius of 32\arcsec; if there were bad pixels or CCD gaps within the region, we shrunk the circle until they were excluded. Therefore, the extraction regions could vary for different observations. We note that the size of the source extraction region does not affect the inferred flux. The background regions were selected to be near the source, off the readout column, and on the same CCD chip as the source. Energy channels of each spectrum were grouped to give a minimum of 25 counts per bin. Spectral fits were performed using XSPEC 12.4 available in HEASOFT 6.4. The errors on the flux and luminosity (absorbed and unabsorbed fluxes) were calculated in XSPEC 12.5 with the {\tt cflux} model. Background subtracted spectra from MOS and PN were fitted simultaneously with the same model in the energy range 0.2--10~keV. Constants to account for possible discrepancy between MOS and PN normalizations were not necessary, thanks to the identical extraction regions and the relative dimness of sources. The absorption column density in our Galaxy along the line of sight to each source was obtained from the LAB map \citep{kal05} using the $N_{\rm H}$ tool on HEASARC, and was set as a lower bound of the total column density in the absorption model. Other parameters specific to each observation are described individually below. All quoted errors in the tables are of the 1$\sigma$ level in order to be consistent with plots. Since the errors on the luminosity are usually smaller than on the temperature and photon index, we treat the luminosity as an independent variable and temperature or photon index as a dependent variable in the curve fitting, and take the errors on the dependent variable only to calculate $\chi^2$.

\subsection{NGC 5204 X-1}

\begin{deluxetable}{cccrrr}[t]
\tablecolumns{6}
\tablewidth{0pc}
\tablecaption{\xmm\ observations of NGC 5204 X-1
\label{tab:ngc5204obs}}

\tablehead{
\colhead{} & \colhead{} & \colhead{} & \multicolumn{3}{c}{Good Exposures (ks)}\\
\cline{4-6}
\colhead{No.} & \colhead{ObsID} & \colhead{Date} & \colhead{PN} & \colhead{MOS1} & \colhead{MOS2}
}
\startdata
 1 & 0142770101 & 2003-01-06 & 15.3 & 18.5 & 18.5 \\
 2 & 0142770301 & 2003-04-25 & 3.8 & 7.6 & 7.6 \\
 3 & 0150650301 & 2003-05-01 & 4.9 & 7.9 & 7.9 \\
 4 & 0405690101 & 2006-11-15 & 9.5 & 17.4 & 17.6 \\
 5 & 0405690201 & 2006-11-19 & 29.0 & 39.7 & 40.7 \\
 6 & 0405690501 & 2006-11-25 & 20.4 & 29.9 & 30.0 \\
\enddata

\tablecomments{Good Exposures are effective exposures after background screening.}
\end{deluxetable}

NGC 5204 X-1 has recently been observed multiple times with \xmm\ and \cha\ \citep{rob05,rob06}. A monitoring program with ten \cha\ snapshots revealed a flux change by a factor of 5 at timescales of a few days \citep{rob06}. By aligning \cha\ and \hst\ images, an optical counterpart to the X-ray source was identified, which appears like an early type supergiant \citep{liu04}. Here, we re-analyzed all six archival \xmm\ observations of the source listed in Table~\ref{tab:ngc5204obs} with observation IDs, dates, and ``good exposures'' that indicate effective exposure times after removing background flares. The first two observations have been reported in the literature \citep[e.g.,][]{rob05,fen05,sto06,win06,vie06}, while the later four have never been published. Our source extraction region has a radius of $27.5\arcsec$ for the first two observations, $20\arcsec$ for the 3rd, and $25\arcsec$ for the last three, respectively. We adopted a distance of 4.3~Mpc to the host galaxy \citep{tul92}, and a Galactic absorption column density of $0.174 \times 10^{21}$~cm$^{-2}$.

\begin{deluxetable*}{ccccccccccc}[t]
\tablecolumns{11}
\tablewidth{0pc}
\tablecaption{Best-fit parameters of NGC 5204 X-1
\label{tab:ngc5204fit}}

\tablehead{
\colhead{No.} & \colhead{$N_{\rm H}$} & \colhead{$\Gamma/\tau$} & \colhead{$N_{\rm PL}/N_{\rm C}$} & \colhead{$T_{\rm e}$} & \colhead{$T_{\rm in}$/$T_0$} & \colhead{$R_{\rm in}\sqrt{\cos i}$} & \colhead{$f_{\rm X}$} & \colhead{$L_{\rm X}$} & \colhead{$L_{\rm bol}$} & \colhead{$\chi^2$/dof}\\
\colhead{(1)} & \colhead{(2)} & \colhead{(3)} & \colhead{(4)} & \colhead{(5)} & \colhead{(6)} & \colhead{(7)} & \colhead{(8)} & \colhead{(9)} & \colhead{(10)} & \colhead{(11)}
}
\startdata
\multicolumn{11}{c}{Model: {\tt wabs$\ast$powerlaw}}\\ \noalign{\smallskip}\hline\noalign{\smallskip}
 1 & $0.59_{-0.03}^{+0.03}$ & $2.07_{-0.02}^{+0.02}$ & $3.58_{-0.06}^{+0.06}$ & \nodata & \nodata & \nodata & $1.62_{-0.02}^{+0.02}$ & $0.430_{-0.005}^{+0.005}$ & \nodata & 513.1/461 \\
 2 & $0.91_{-0.06}^{+0.07}$ & $2.29_{-0.04}^{+0.04}$ & $5.60_{-0.17}^{+0.17}$ & \nodata & \nodata & \nodata & $2.02_{-0.04}^{+0.04}$ & $0.619_{-0.014}^{+0.015}$ & \nodata & 273.8/219 \\
 3 & $1.13_{-0.06}^{+0.06}$ & $2.42_{-0.04}^{+0.04}$ & $7.06_{-0.19}^{+0.20}$ & \nodata & \nodata & \nodata & $2.24_{-0.04}^{+0.04}$ & $0.761_{-0.017}^{+0.019}$ & \nodata & 288.3/262 \\
 4 & $1.41_{-0.04}^{+0.04}$ & $2.66_{-0.02}^{+0.02}$ & $10.06_{-0.18}^{+0.18}$ & \nodata & \nodata & \nodata & $2.64_{-0.03}^{+0.03}$ & $1.08_{-0.02}^{+0.02}$ & \nodata & 593.5/566 \\
 5 & $1.40_{-0.03}^{+0.03}$ & $2.561_{-0.016}^{+0.016}$ & $8.35_{-0.10}^{+0.10}$ & \nodata & \nodata & \nodata & $2.298_{-0.015}^{+0.015}$ & $0.892_{-0.011}^{+0.011}$ & \nodata & 926.8/817 \\
 6 & $0.83_{-0.03}^{+0.03}$ & $2.251_{-0.018}^{+0.020}$ & $4.95_{-0.07}^{+0.07}$ & \nodata & \nodata & \nodata & $1.866_{-0.018}^{+0.018}$ & $0.553_{-0.006}^{+0.006}$ & \nodata & 807.0/658 \\
\cutinhead{Model: {\tt wabs(powerlaw + diskbb)}}
 1 & $0.56_{-0.05}^{+0.05}$ & $1.91_{-0.05}^{+0.04}$ & $2.93_{-0.19}^{+0.16}$ & \nodata & $0.26_{-0.02}^{+0.03}$ & $0.75_{-0.15}^{+0.16}$ & $1.65_{-0.02}^{+0.02}$ & $0.430_{-0.008}^{+0.008}$ & \nodata & 491.8/459 \\
 2 & $0.71_{-0.10}^{+0.14}$ & $1.92_{-0.10}^{+0.10}$ & $3.5_{-0.5}^{+0.5}$ & \nodata & $0.29_{-0.03}^{+0.03}$ & $1.01_{-0.18}^{+0.33}$ & $2.11_{-0.04}^{+0.04}$ & $0.59_{-0.02}^{+0.03}$ & \nodata & 249.1/217 \\
 3 & $0.82_{-0.09}^{+0.15}$ & $2.10_{-0.10}^{+0.12}$ & $4.6_{-0.6}^{+0.9}$ & \nodata & $0.33_{-0.05}^{+0.03}$ & $0.72_{-0.12}^{+0.24}$ & $2.30_{-0.04}^{+0.04}$ & $0.68_{-0.03}^{+0.03}$ & \nodata & 277.1/260 \\
 4 & $1.06_{-0.07}^{+0.12}$ & $2.42_{-0.06}^{+0.08}$ & $7.0_{-0.6}^{+0.9}$ & \nodata & $0.36_{-0.03}^{+0.02}$ & $0.64_{-0.11}^{+0.12}$ & $2.66_{-0.03}^{+0.03}$ & $0.90_{-0.04}^{+0.04}$ & \nodata & 578.3/564 \\
 5 & $0.94_{-0.04}^{+0.07}$ & $2.18_{-0.04}^{+0.05}$ & $4.8_{-0.3}^{+0.4}$ & \nodata & $0.352_{-0.016}^{+0.009}$ & $0.73_{-0.04}^{+0.06}$ & $2.348_{-0.017}^{+0.017}$ & $0.722_{-0.016}^{+0.017}$ & \nodata & 837.0/815 \\
 6 & $0.63_{-0.06}^{+0.04}$ & $1.85_{-0.06}^{+0.04}$ & $2.96_{-0.25}^{+0.18}$ & \nodata & $0.297_{-0.010}^{+0.018}$ & $0.90_{-0.10}^{+0.07}$ & $1.96_{-0.02}^{+0.02}$ & $0.522_{-0.008}^{+0.009}$ & \nodata & 697.7/656 \\
\cutinhead{Model: {\tt wabs$\ast$comptt}}
 1 & $0.17^{<0.21}$ & $5.61_{-0.54}^{+0.08}$ & $4.0_{-0.9}^{+0.3}$ & $2.88_{-0.23}^{+0.19}$ & $0.116_{-0.005}^{+0.003}$ & \nodata & $1.61_{-0.02}^{+0.02}$ & $0.374_{-0.004}^{+0.005}$ & 0.44 & 492.7/459 \\
 2 & $0.17^{<0.25}$ & $0.151_{-0.009}^{+0.012}$ & $0.142_{-0.003}^{+0.012}$ & [100] & $0.141_{-0.005}^{+0.003}$ & \nodata & $2.04_{-0.04}^{+0.04}$ & $0.471_{-0.008}^{+0.022}$ & 0.67 & 243.2/218 \\
 3 & $0.29_{-0.07}^{+0.14}$ & $0.125_{-0.007}^{+0.009}$ & $0.172_{-0.016}^{+0.024}$ & [100] & $0.142_{-0.003}^{+0.005}$ & \nodata & $2.28_{-0.04}^{+0.04}$ & $0.55_{-0.02}^{+0.02}$ & 0.73 & 275.9/261 \\
 4 & $0.26_{-0.07}^{+0.08}$ & $4.8_{-0.7}^{+0.6}$ & $9._{-2.}^{+2.}$ & $2.4_{-2.4}^{+0.7}$ & $0.152_{-0.008}^{+0.007}$ & \nodata & $2.62_{-0.03}^{+0.03}$ & $0.628_{-0.016}^{+0.017}$ & 0.67 & 570.5/564 \\
 5 & $0.193_{-0.014}^{+0.014}$ & $0.0983_{-0.0031}^{+0.0017}$ & $0.160_{-0.005}^{+0.005}$ & [100] & $0.1647_{-0.0013}^{+0.0014}$ & \nodata & $2.326_{-0.015}^{+0.015}$ & $0.542_{-0.008}^{+0.008}$ & 0.68 & 828.7/816 \\
 6 & $0.17^{<0.21}$ & $0.167_{-0.005}^{+0.007}$ & $0.1315_{-0.0012}^{+0.0065}$ & [100] & $0.1359_{-0.0032}^{+0.0016}$ & \nodata & $1.889_{-0.015}^{+0.021}$ & $0.438_{-0.003}^{+0.008}$ & 0.64 & 728.4/657 \\
\cutinhead{Model: {\tt wabs(comptt + diskbb)}}
 1 & $0.29_{-0.05}^{+0.05}$ & $10.3_{-1.2}^{+1.8}$ & $2.5_{-0.3}^{+0.3}$ & $1.59_{-0.15}^{+0.17}$ & $0.27_{-0.03}^{+0.03}$ & $1.2_{-0.2}^{+0.2}$ & $1.61_{-0.02}^{+0.02}$ & $0.385_{-0.007}^{+0.008}$ & 0.45 & 462.8/458 \\
 2 & $0.51_{-0.10}^{+0.11}$ & $0.20_{-0.03}^{+0.05}$ & $0.082_{-0.015}^{+0.021}$ & [100] & $0.209_{-0.010}^{+0.017}$ & $2.2_{-0.3}^{+0.5}$ & $2.08_{-0.04}^{+0.05}$ & $0.54_{-0.02}^{+0.02}$ & 0.87 & 242.9/217 \\
 3 & $0.50_{-0.11}^{+0.10}$ & $0.132_{-0.015}^{+0.024}$ & $0.12_{-0.03}^{+0.05}$ & [100] & $0.188_{-0.039}^{+0.019}$ & $2.4_{-0.3}^{+0.5}$ & $2.28_{-0.04}^{+0.04}$ & $0.59_{-0.03}^{+0.02}$ & 0.85 & 274.1/260 \\
 4 & $0.52_{-0.07}^{+0.06}$ & $5.2_{-1.1}^{+0.8}$ & $6.7_{-1.9}^{+1.6}$ & $2.2_{-0.4}^{+0.9}$ & $0.20_{-0.02}^{+0.02}$ & $2.6_{-0.3}^{+0.3}$ & $2.62_{-0.03}^{+0.03}$ & $0.688_{-0.018}^{+0.018}$ & 0.83 & 559.2/563 \\
 5 & $0.56_{-0.03}^{+0.03}$ & $4.8_{-2.1}^{+1.1}$ & $2.8_{-1.8}^{+0.6}$ & $2.9_{-0.7}^{+2.0}$ & $0.253_{-0.015}^{+0.014}$ & $1.72_{-0.17}^{+0.23}$ & $2.322_{-0.022}^{+0.017}$ & $0.610_{-0.008}^{+0.008}$ & 0.75 & 803.6/814 \\
 6 & $0.44_{-0.04}^{+0.04}$ & $0.28_{-0.03}^{+0.03}$ & $0.059_{-0.004}^{+0.006}$ & [100] & $0.244_{-0.003}^{+0.003}$ & $1.58_{-0.07}^{+0.11}$ & $1.95_{-0.02}^{+0.02}$ & $0.489_{-0.007}^{+0.008}$ & 0.89 & 690.7/656 \\
\enddata

\tablecomments{
Col.~(1): Observation index corresponding to Column (1) of Table~\ref{tab:ngc5204obs}.
Col.~(2): Absorption column density in units of $10^{21}$~cm$^{-2}$.
Col.~(3): $\Gamma$ is the photon index of the {\tt powerlaw} model; $\tau$ is the optical depth of the {\tt comptt} model.
Col.~(4): $N_{\rm PL}$ is the normalization of the {\tt powerlaw} model at 1~keV in units of $10^{-4}$~photons~cm$^{-2}$~s$^{-1}$; $N_{\rm C}$ is the normalization of the {\tt comptt} model in units of $10^{-4}$.
Col.~(5): Plasma temperature in units of keV; values in box brackets are fixed in the fitting.
Col.~(6): Inner disk temperature $T_{\rm in}$ of the {\tt diskbb} model or seed photon temperature $T_0$ of the {\tt comptt} model in units of keV. For {\tt comptt + diskbb}, $T_0$ is set equal to $T_{\rm in}$.
Col.~(7): $R_{\rm in}$ is the inner disk radius in units of $10^3$~km; $i$ is the disk inclination angle.
Col.~(8): Absorbed flux in 0.3--10 keV in units of $10^{-12}$~\ergcms.
Col.~(9): Unabsorbed luminosity in 0.3--10 keV in units of $10^{40}$~\ergs.
Col.~(10): bolometric luminosity (integration of physical models in 0.01--100 keV) in units of $10^{40}$~\ergs.
Col.~(11): Best-fit $\chi^2$ and degrees of freedom.
All errors are at 1 $\sigma$ level.
}
\end{deluxetable*}

\begin{figure*}[t]
\centering
\includegraphics[width=0.49\textwidth]{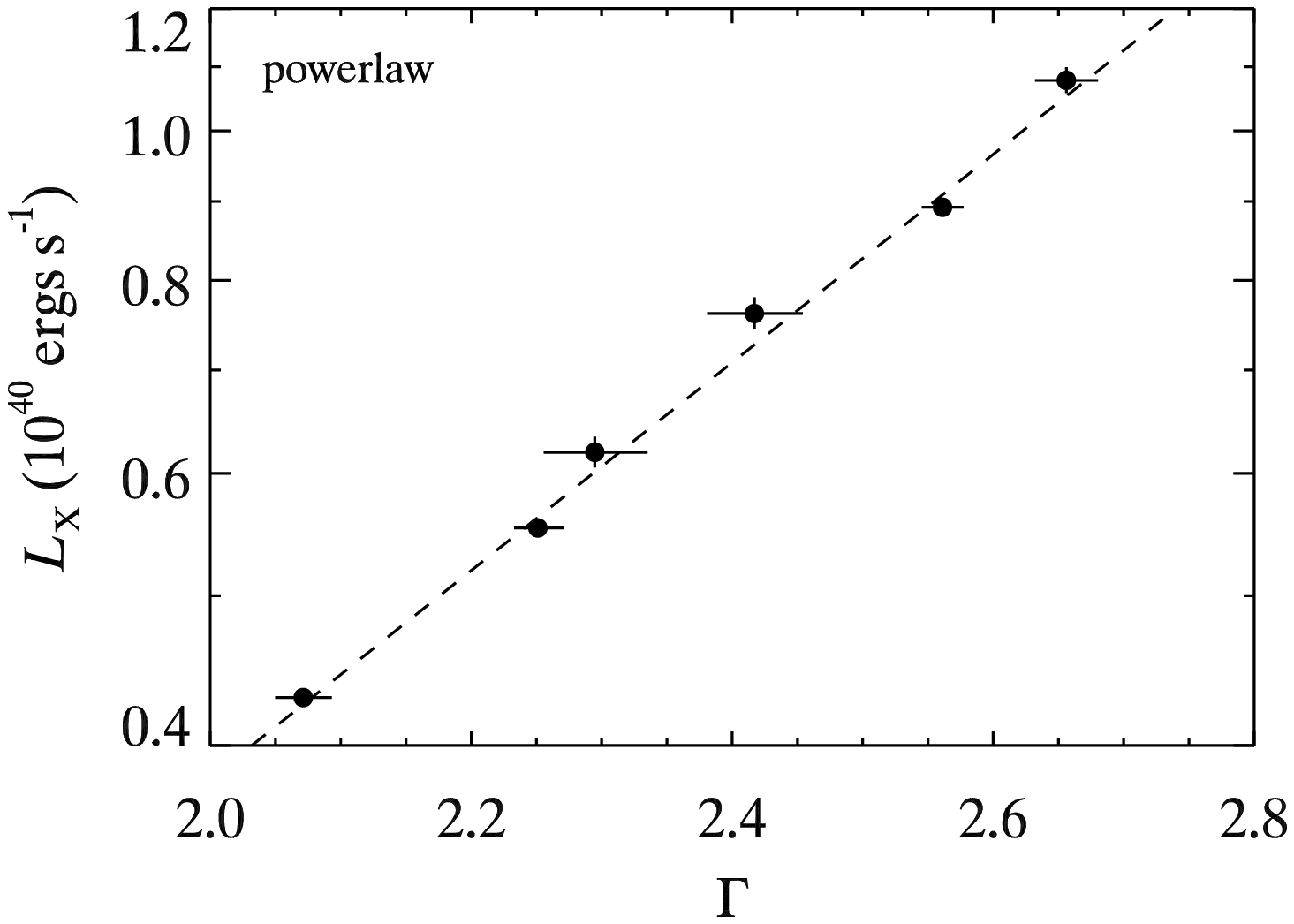}
\includegraphics[width=0.49\textwidth]{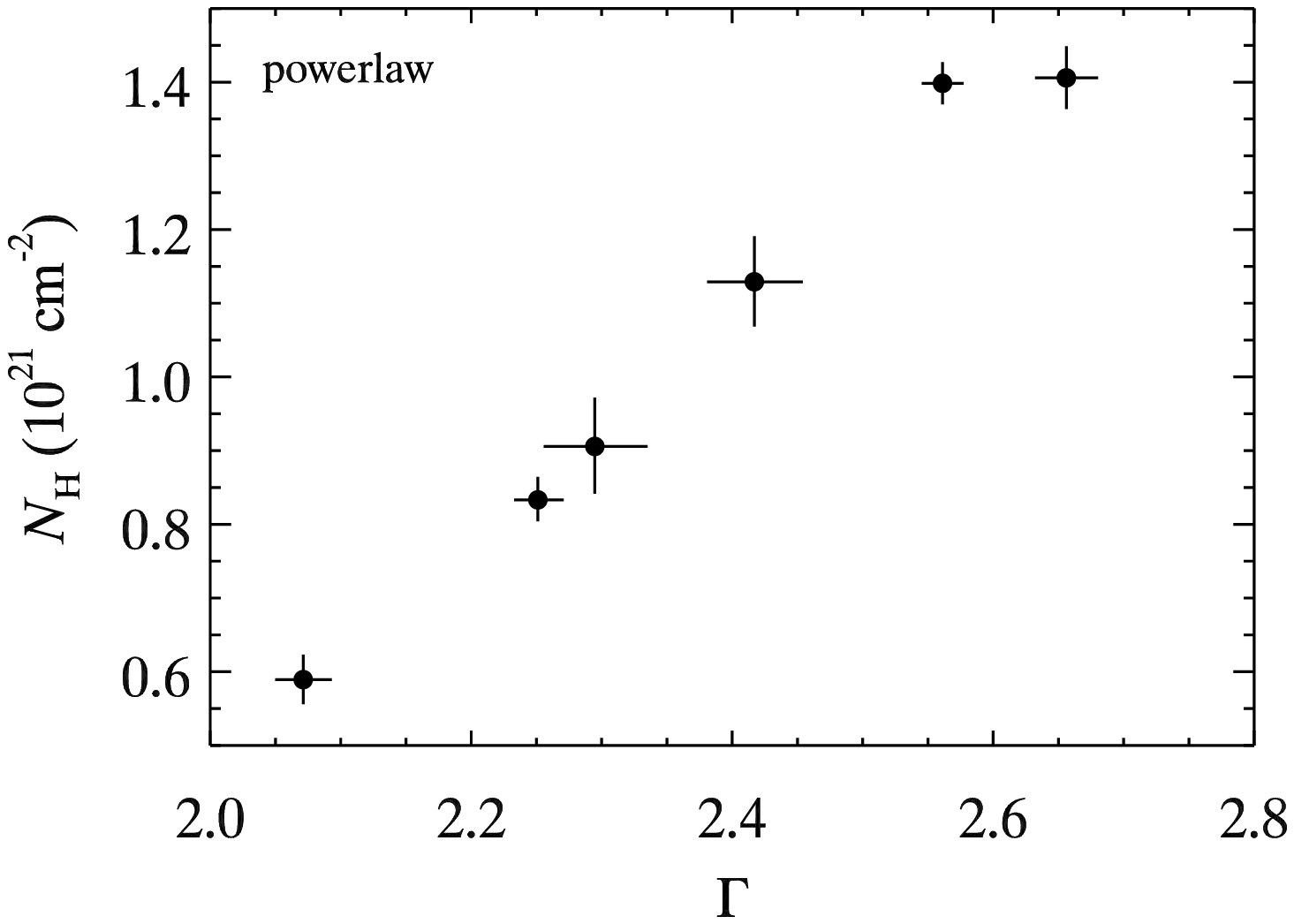}
\includegraphics[width=0.49\textwidth]{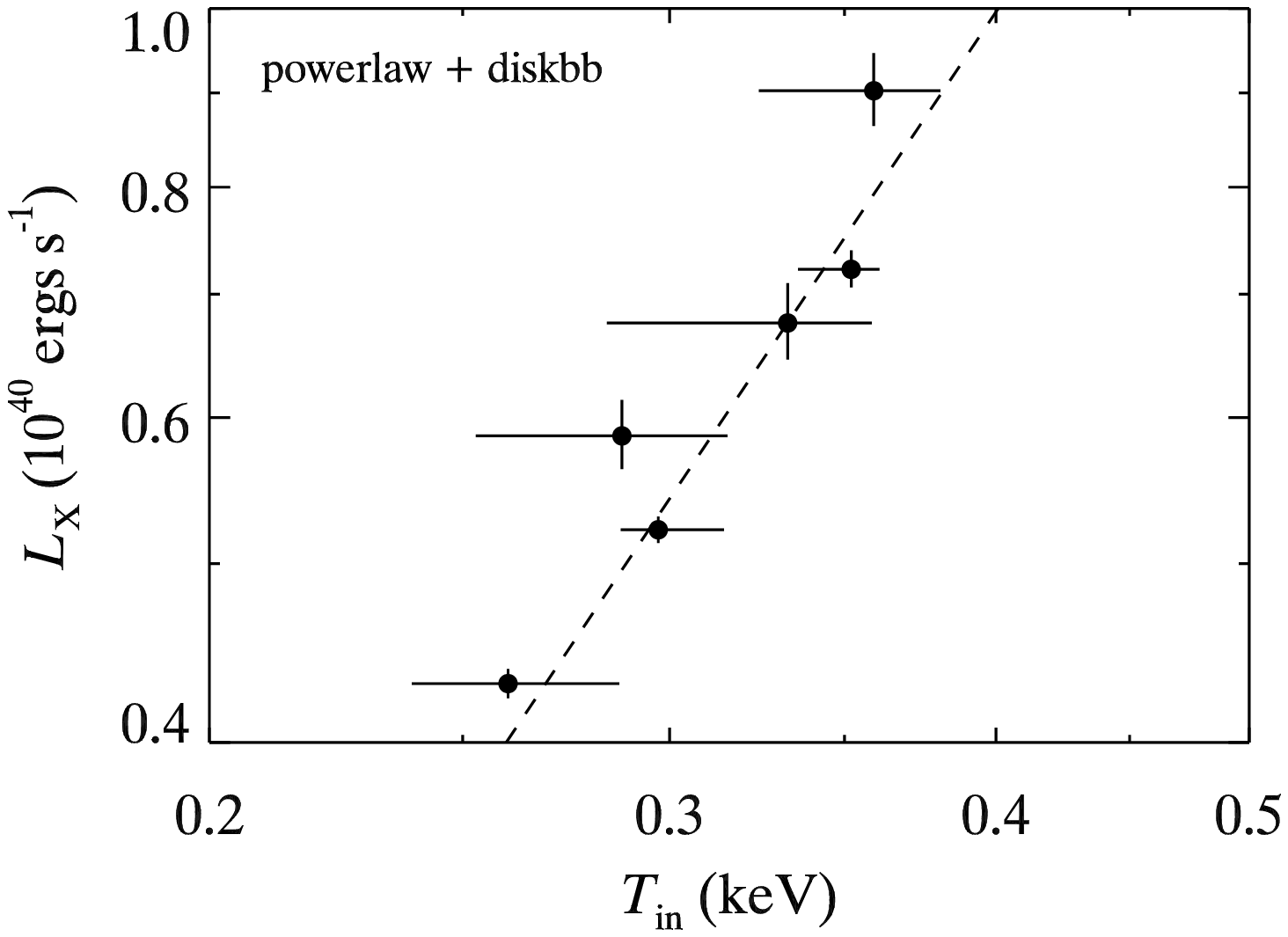}
\includegraphics[width=0.49\textwidth]{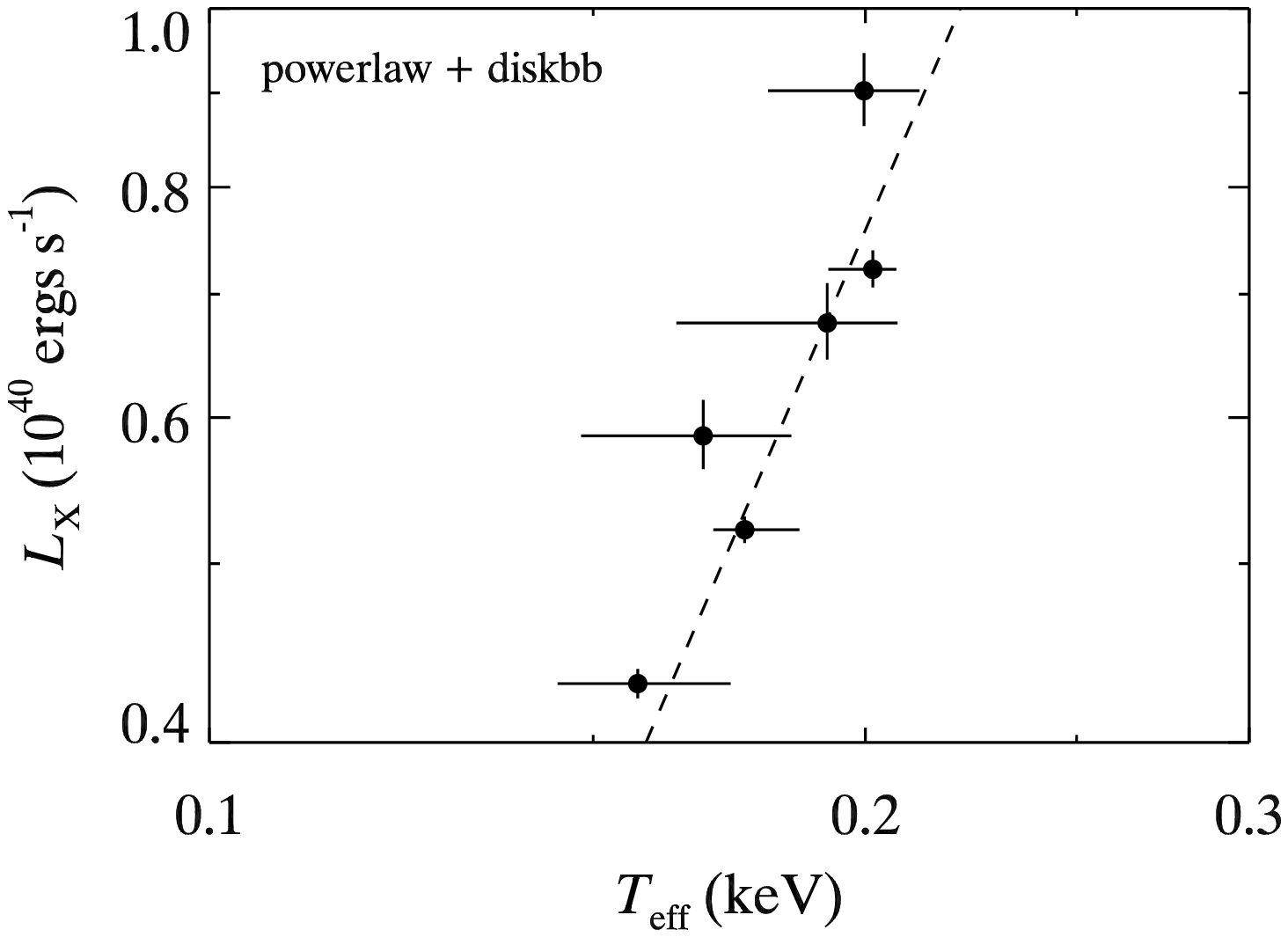}
\includegraphics[width=0.49\textwidth]{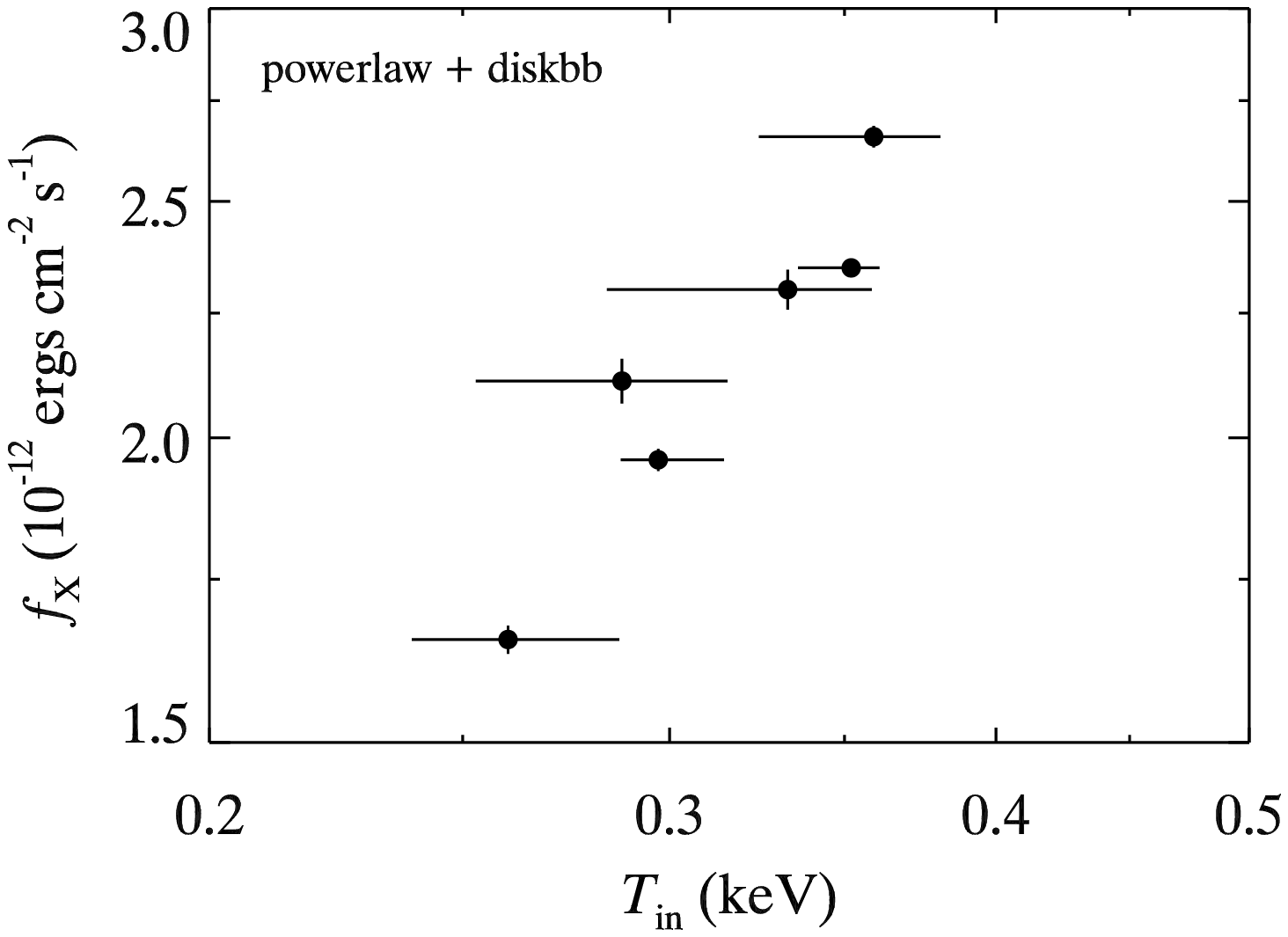}
\includegraphics[width=0.49\textwidth]{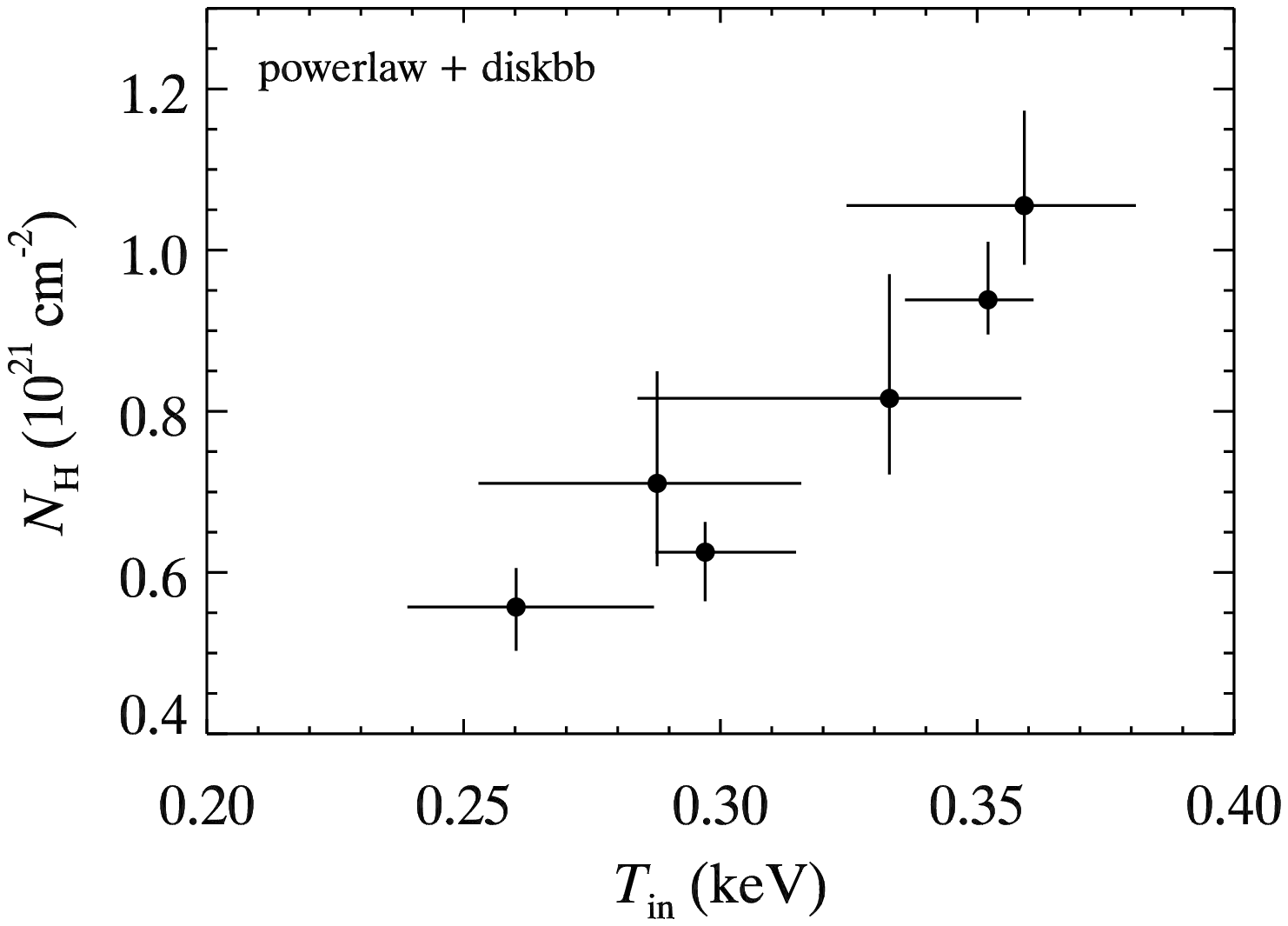}
\caption{
Diagrams between best-fit spectral parameters of NGC 5204 X-1 derived from different models (see Table~\ref{tab:ngc5204fit}). $T_{\rm eff}$ is calculated from $T_{\rm in}$ with hardening correction (see texts for details). Dashed lines indicate best-fit relations: $\log(L_{\rm X}) \propto (0.67 \pm 0.03)\Gamma$, $L_{\rm X} \propto T_{\rm in}^{2.1 \pm 0.5}$, and $L_{\rm X} \propto T_{\rm eff}^{3.7 \pm 1.6}$. The correlation coefficient $r$ and chance probability $p$ are: $r=0.995$, $p=3.3\times10^{-5}$ between $\log L_{\rm X}$ and $\Gamma$, $r=0.95$, $p=3.5 \times 10^{-3}$ between $\log L_{\rm X}$ and $\log T_{\rm in}$, and $r=0.92$, $p=9.8 \times 10^{-3}$ between $\log L_{\rm X}$ and $\log T_{\rm eff}$.
\label{fig:ngc5204}}
\end{figure*}

\addtocounter{figure}{-1}
\begin{figure*}
\centering
\includegraphics[width=0.49\textwidth]{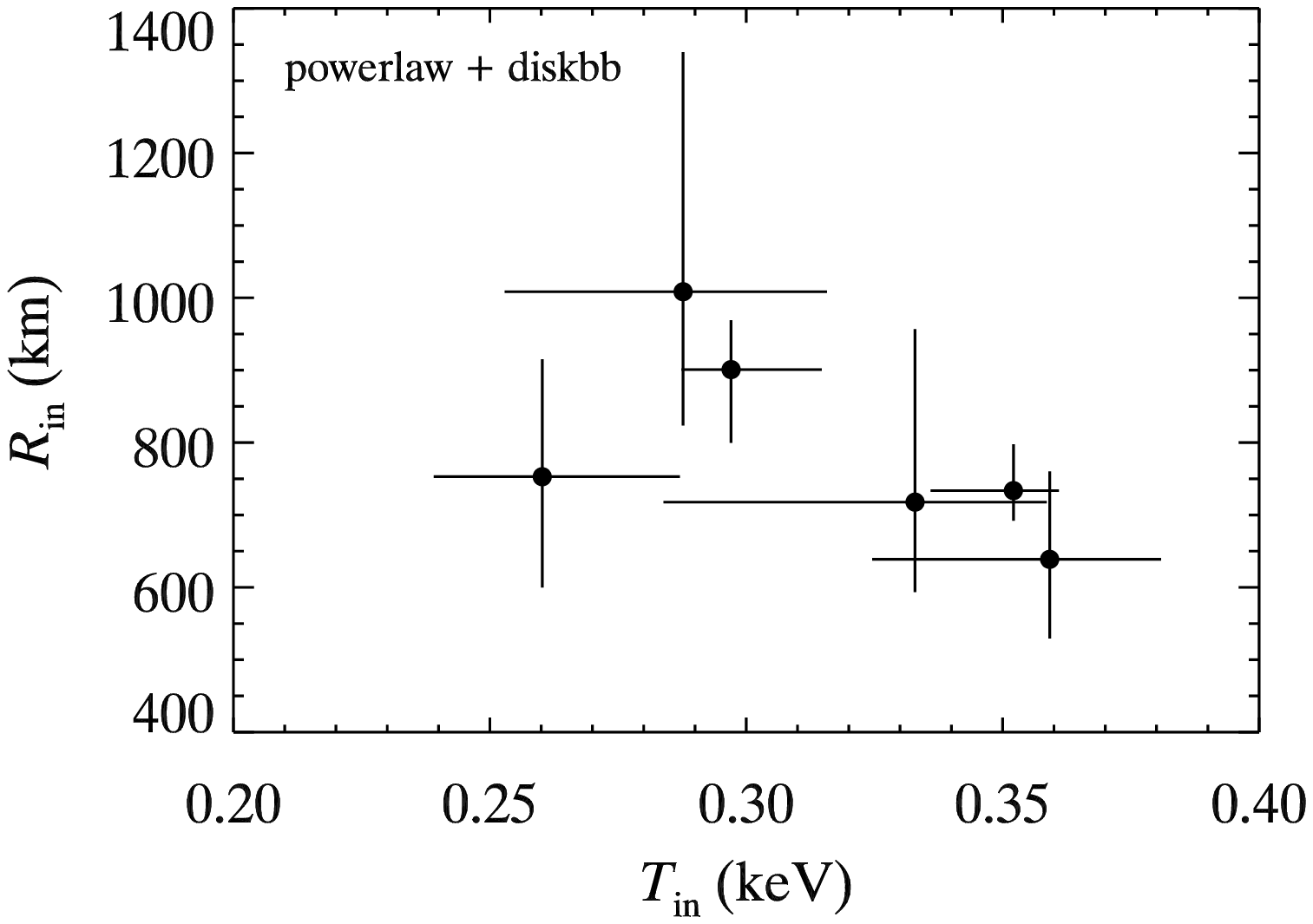}
\includegraphics[width=0.49\textwidth]{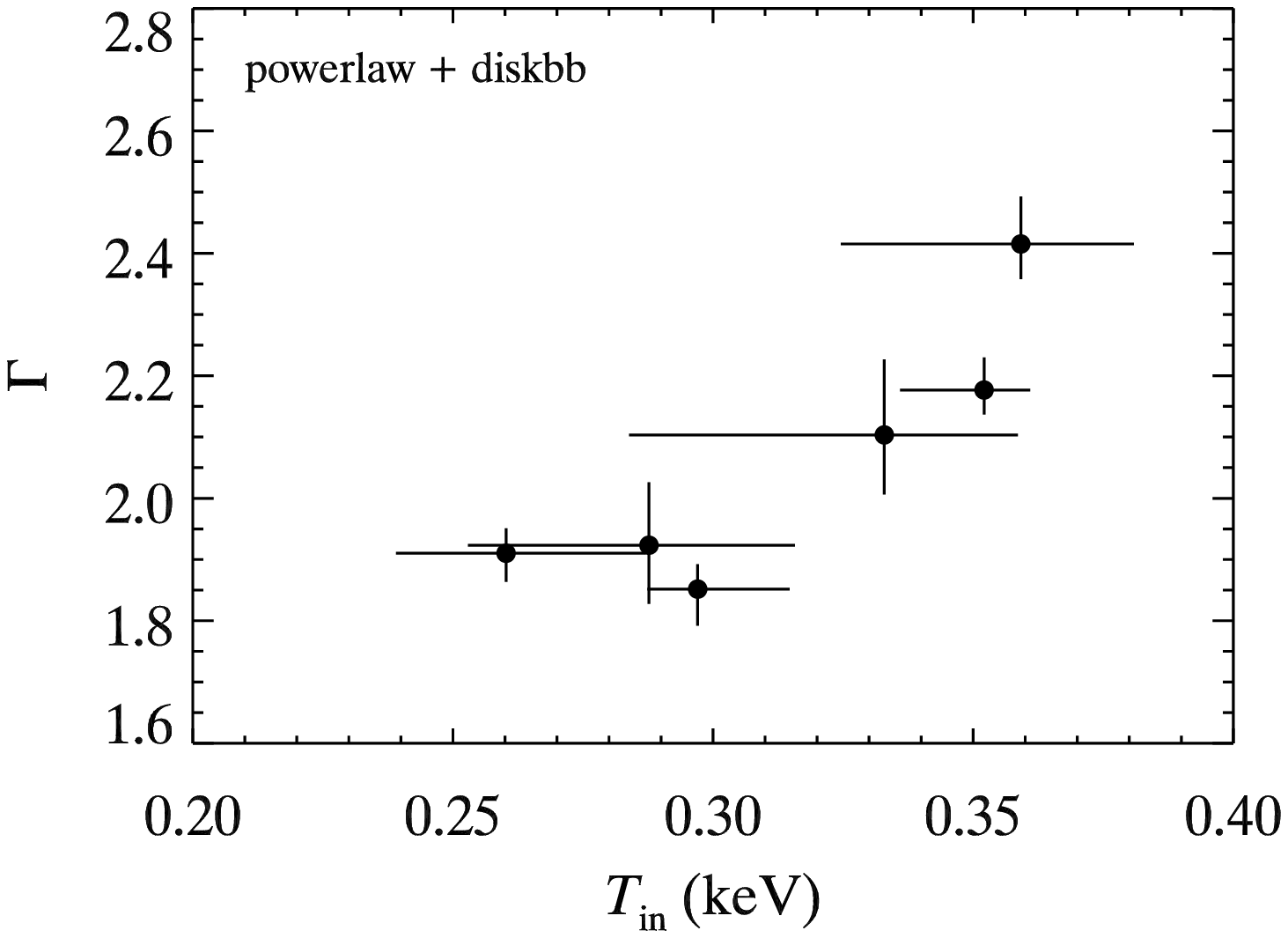}
\includegraphics[width=0.49\textwidth]{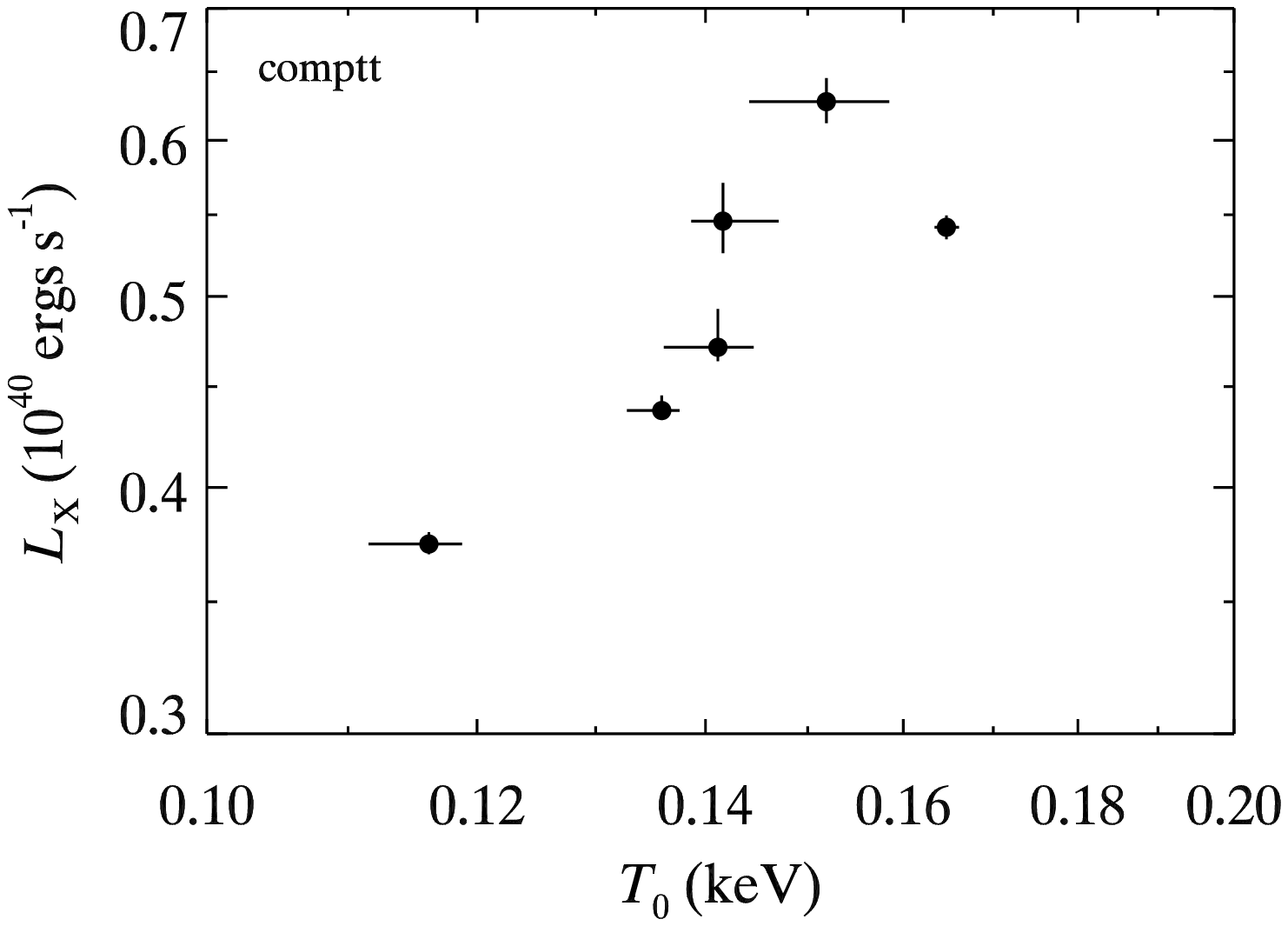}
\includegraphics[width=0.49\textwidth]{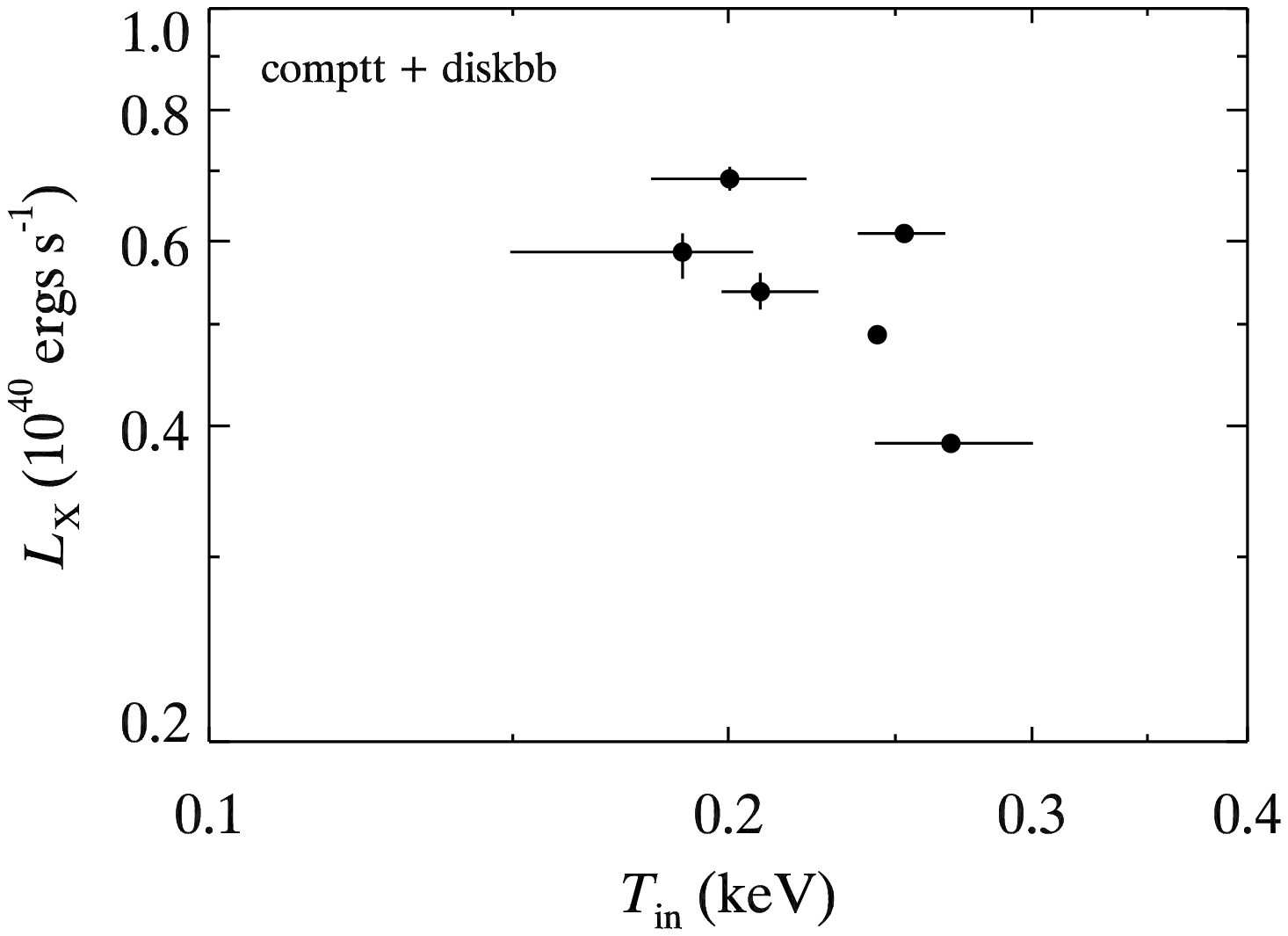}
\caption{Continued.}
\end{figure*}

We tried different models to fit the data. All spectral parameters, observed fluxes $f_{\rm X}$ and unabsorbed luminosities $L_{\rm X}$ in the 0.3--10 keV band, $\chi^2$, and degrees of freedoms are listed in Table~\ref{tab:ngc5204fit}. If all model components are physical (with a finite integration), a bolometric luminosity $L_{\rm bol}$ is also calculated by integrating the model (without absorption) in the energy range of 0.01--100 keV. First, we tried an absorbed power-law model ({\tt powerlaw} in XSPEC), as the first-order approximation of the spectral shape. A cool disk emission component ({\tt diskbb} in XSPEC) was then added, which improved the fits significantly. To replace the power-law component with a physical model, we tried a Comptonization model ({\tt comptt} in XSPEC), with and without a disk component. When the disk component is added, the disk inner temperature $T_{\rm in}$ and the Comptonization seed photon temperature $T_0$ are equalized in the fits, i.e., assuming that the disk emission provides source photons for Comptonization. In some observations, the spectra show a featureless power-law tail at energies above 2 or 3~keV, leading to difficulty in finding the electron temperature $T_{\rm e}$, which is determined by the position of the high energy bend. In those cases, we fixed $T_{\rm e}$ at 100~keV to create a featureless power-law tail below 10~keV; higher or slightly lower temperatures do not affect other spectral parameters and the goodness of fit.

Diagrams between pairs of parameters of interest are plotted in Figure~\ref{fig:ngc5204}. The model being used to derive the parameters of that plot is indicated on the upper left corner of the panel (not including the absorption component). A tight, linear correlation was found between $\log L_{\rm X}$ and the power-law photon index $\Gamma$ when fitting the data with an absorbed power-law model. The best-fit relation is $\log L_{\rm X} \propto (0.67 \pm 0.03) \Gamma$.  The correlation coefficient between $\log L_{\rm X}$ and $\Gamma$ is 0.995 with a chance probability of $3.3\times10^{-5}$. The absorption column density $N_{\rm H}$ is also correlated with $\Gamma$, and seems to saturate when $\Gamma \ga 2.6$.

When an additional disk component is added to account for soft excesses, a correlation appears between $\log L_{\rm X}$ and $\log T_{\rm in}$, with a coefficient of 0.95 and a chance probability of $3.5 \times 10^{-3}$. Fitting with a power-law relation, we derive $L_{\rm X} \propto T_{\rm in}^{2.1 \pm 0.5}$. This resembles the intrinsic evolution expected for a disk blackbody, except that the exponent of the relation is lower than 4. $T_{\rm in}$ derived from the model {\tt diskbb} is the color temperature, while the $L \propto T^4$ relation is actually expected between the bolometric luminosity and effective temperature $T_{\rm eff}$ \citep{mcc08}. We thus calculated hardening correction $f \equiv T_{\rm in}/T_{\rm eff}$ as a function of luminosity in Eddington units by comparing the disk atmosphere model {\tt bhspec} \citep{dav05,dav06} with the {\tt kerrbb} model \citep{lil05} assuming a non-spinning 100 $M_\sun$ black hole (fitting with {\tt kerrbb} to spectra generated based on {\tt bhspec}), and then computed $T_{\rm eff}$ assuming the highest observed luminosity is the Eddington limit. The computed $f$ goes almost linearly from 1.5 to 2.1 in the luminosity range of 0.1 to 1 Eddington limit (see \citet{mcc08} for a similar plot). After applying the hardening correction, we obtained $L_{\rm X} \propto T_{\rm eff}^{3.7 \pm 1.6}$. The exponent of the relation between $L_{\rm X}$ and $T_{\rm eff}$ decreases if the hardening correction is computed at a lower Eddington level, e.g., we find an exponent of $2.7 \pm 0.9$ if assuming the highest observed luminosity is half of the Eddington limit. This hardening correction is model dependent, but the basic behavior of the luminosity versus temperature is similar to that found for stellar-mass black hole X-ray binaries.  The disk inner radius derived from the {\tt diskbb} model is about 1000~km, with a tendency to decrease at high temperatures. A correlation also exists between $f_{\rm X}$ and $T_{\rm in}$, which strengthens confidence in the relation between $L_{\rm X}$ and $T_{\rm in}$. $N_{\rm H}$ goes higher with $T_{\rm in}$ and $L_{\rm X}$, indicative of a denser environment when the accretion rate increases. The power-law component seems to be related to the thermal component, shown as a correlation between $\Gamma$ and $T_{\rm in}$. The disk emission contributes 0.11, 0.22, 0.19, 0.15, 0.23, and 0.23 of the total luminosity in the 0.3--10~keV band, respectively for the six observations with a disk plus power-law model.

\begin{figure}[t]
\centering
\includegraphics[width=\columnwidth]{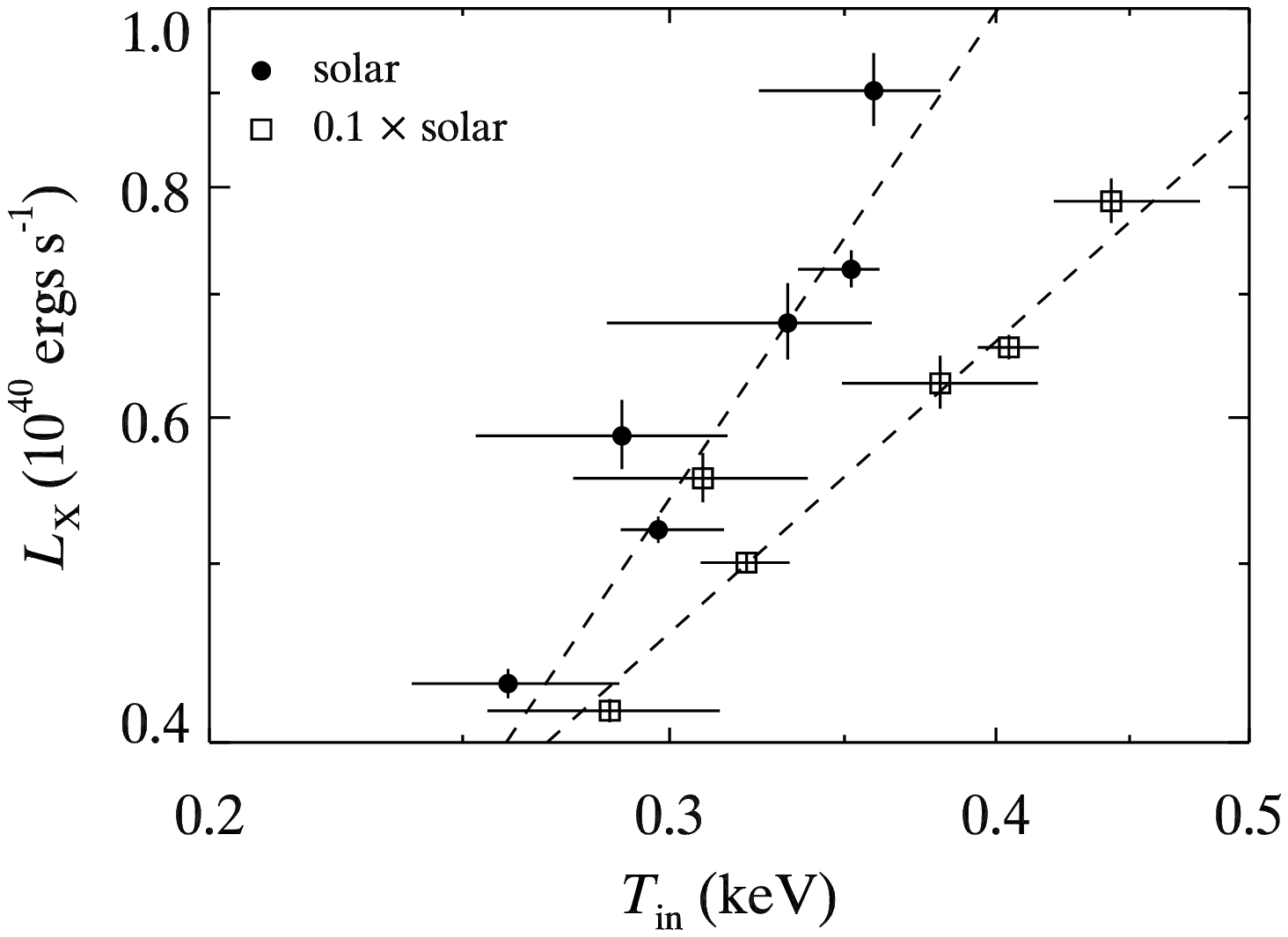}
\caption{
$L_{\rm X}$ versus $T_{\rm in}$ inferred from the {\tt diskbb} plus {\tt powerlaw} model with absorption in 1 and 0.1 solar abundances in the host galaxy, respectively. Data points for 1 solar abundance are the same as in Figure~\ref{fig:ngc5204}. With 0.1 solar abundance, the best-fit relation is $L \propto T_{\rm in}^{1.3 \pm 0.2}$, and the correlation coefficient between $\log L_{\rm X}$ and $\log T_{\rm in}$ is 0.97 with a chance probability of $1.1 \times 10^{-3}$. 
\label{fig:ngc5204var}}
\end{figure}

The cool disk component peaks at energies where the absorption takes significant effect. Therefore, how to model the absorption must be considered carefully. We tried {\tt tbabs} instead of {\tt wabs} in XSPEC but no significant changes were found. We then tested effects caused by different abundances. \citet{win07} found that most ULXs including this one have near-solar abundances by measuring oxygen absorption edge in \xmm\ spectra. \citet{rob06} found sub-solar abundance could marginally improve the fits to \cha\ data of this source. The source is also found near young star clusters \citep{goa02}. These results suggest that the abundance is close to or, perhaps, somewhat lower than solar. To see the effect of abundance, we modeled absorption within the Milky Way using {\tt wabs} with $N_{\rm H}$ fixed at the Galactic value and the abundance set to solar and absorption within the host galaxy using {\tt vphabs} with free $N_{\rm H}$ and the abundance set to 0.1 solar. The model consists of an accretion disk and a power-law. With sub-solar abundance, the derived luminosity is lower and the disk temperature is higher than with solar abundance (see Figure~\ref{fig:ngc5204var}). Fitting with a power-law relation, we obtain $L_{\rm X} \propto T_{\rm in}^{1.3 \pm 0.2}$, and a correlation coefficient of 0.97 between $\log L_{\rm X}$ and $\log T_{\rm in}$ with a chance probability of $1.1 \times 10^{-3}$. Other abundances from 0.1 to 1 solar value will produce a slope in between. Therefore, the correlation between $\log L_{\rm X}$ and $\log T_{\rm in}$ appears robust even allowing for the uncertainty in the abundance of the host galaxy. 

If we model the spectra using an absorbed Comptonization model, the derived absorption column density is rather low, often pegged at the Galactic value, which means no extra absorption is found in the host galaxy or internal to the binary system. This is somewhat unphysical and hard to explain. In observations 1 and 4, the best-fit $T_{\rm e}$ is lower than 3~keV, as a result of spectral curvature. The optical depth in these two observations is also high, indicative of a cool, optically thick electron corona dominating the emission. However, the source in these two observations did not show differences on flux and absorption from the other four, in which it had power-law spectral tails implying emission from a hot, optically thin corona. The luminosity derived from the absorbed Comptonization model presents a positive correlation with the seed photon temperature, with a relation of $L_{\rm X} \propto T_0^{1.27 \pm 0.09}$ or $L_{\rm bol} \propto T_0^{1.43 \pm 0.15}$. Then, we added a disk component in order to account for non Comptonized disk photons. With this model, a cool corona at low significance is also shown in observation 5; if we fix $T_{\rm e}$ at 100~keV, the $\chi^2$ only increases 1.7 and no significant changes are observed on $N_{\rm H}$ and $T_0$, but the optical depth $\tau$ becomes 0.11 due to the degeneracy between $\tau$ and $T_{\rm e}$. The luminosity, $L_{\rm X}$, $L_{\rm bol}$ or $L_{\rm disk}$, becomes small when $T_{\rm in}$ goes up, which is obviously in conflict with the $L \propto T^4$ relation. 

\begin{figure}[t]
\centering
\includegraphics[width=0.8\columnwidth]{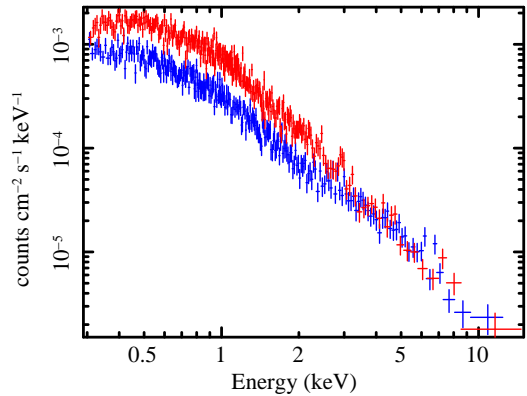}
\caption{
\xmm\ PN spectra of NGC 5204 X-1 in observations 4 (red) and 1 (blue). It is clearly seen that the high state spectrum has excesses over the low one at energies up to 3~keV, where absorption takes very little effect. The low state spectrum cannot be modeled as the spectrum from the high state with extra absorption. In the energy band above 3~keV, the two spectra have almost the same flux level, indicating the flux variation is due to a soft emission component.
\label{fig:ngc5204hl}}
\end{figure}

The disk blackbody plus power-law model is preferred for NGC 5204 X-1, because a correlation with moderate confidence level was found and a self-consistent interpretation can be made. However, replacing the power-law component with a physical (Comptonization) model fails to produce the anticipated correlation. This is probably because that the corona hides the innermost part of the disk when it is optically thick, and $T_{\rm in} = T_0$ is no longer valid in this case. Disconnecting $T_{\rm in}$ and $T_0$ would lead to degeneracy in the model fits, making it difficult or impossible to find a unique global minimum, especially since the Comptonization model alone provides adequate fits. A real physical model that includes the disk emission and Comptonization self consistently could be evaluated when hard X-ray data are available and tight constraints on the Comptonization parameters can be obtained.

In Figure~\ref{fig:ngc5204hl}, two background subtracted energy spectra were plotted together, one from observation 4 with the highest flux among all, and the other from observation 1 with the lowest flux. The two spectra were normalized in units of counts~cm$^{-2}$~s$^{-1}$~keV$^{-1}$ to account for different exposures and extraction areas. It is clear that the high flux spectrum is above the low one at energies below 3~keV, and they are almost at same levels above 3~keV. This means that the flux variation is from a soft emission component, and is not caused by absorption that takes almost no effect at energy around 2--3 keV with $N_{\rm H} \la 10^{21}$~cm$^{-2}$.

\subsection{Holmberg II X-1}

\begin{deluxetable}{cccrrr}[t]
\tablecolumns{6}
\tablewidth{0pc}
\tablecaption{\xmm\ observations of Holmberg II X-1
\label{tab:hoiiobs}}

\tablehead{
\colhead{} & \colhead{} & \colhead{} & \multicolumn{3}{c}{Good Exposures (ks)}\\
\cline{4-6}
\colhead{No.} & \colhead{ObsID} & \colhead{Date} & \colhead{PN} & \colhead{MOS1} & \colhead{MOS2}
}
\startdata
 1 & 0112520601 & 2002-04-10 & 4.6 & 10.1 & 10.1 \\
 2 & 0112520701 & 2002-04-16 & 3.5 & 5.3 & 5.5 \\
 3 & 0112520901 & 2002-09-18 & 4.2 & 6.5 & 6.5 \\
 4 & 0200470101 & 2004-04-15 & 35.0 & \nodata  & \nodata \\
\enddata

\tablecomments{Good Exposures are effective exposures after background screening. MOS data are not used for observation 4.}
\end{deluxetable}

\begin{deluxetable*}{ccccccccccc}[t]
\tablecolumns{11}
\tablewidth{0pc}
\tablecaption{Best-fit parameters of Holmberg II X-1
\label{tab:hoiifit}}

\tablehead{
\colhead{No.} & \colhead{$N_{\rm H}$} & \colhead{$\Gamma/\tau$} & \colhead{$N_{\rm PL}/N_{\rm C}$} & \colhead{$T_{\rm e}$} & \colhead{$T_{\rm in}$/$T_0$} & \colhead{$R_{\rm in}\sqrt{\cos i}$} & \colhead{$f_{\rm X}$} & \colhead{$L_{\rm X}$} & \colhead{$L_{\rm bol}$} & \colhead{$\chi^2$/dof}\\
\colhead{(1)} & \colhead{(2)} & \colhead{(3)} & \colhead{(4)} & \colhead{(5)} & \colhead{(6)} & \colhead{(7)} & \colhead{(8)} & \colhead{(9)} & \colhead{(10)} & \colhead{(11)}
}
\startdata
\multicolumn{11}{c}{Model: {\tt wabs$\ast$powerlaw}}\\ \noalign{\smallskip}\hline\noalign{\smallskip}
 1 & $1.67_{-0.04}^{+0.04}$ & $2.70_{-0.02}^{+0.02}$ & $27.4_{-0.4}^{+0.4}$ & \nodata & \nodata & \nodata & $6.56_{-0.05}^{+0.05}$ & $1.83_{-0.03}^{+0.03}$ & \nodata & 682.2/653 \\
 2 & $1.33_{-0.04}^{+0.04}$ & $2.47_{-0.03}^{+0.03}$ & $21.0_{-0.4}^{+0.4}$ & \nodata & \nodata & \nodata & $6.20_{-0.07}^{+0.07}$ & $1.40_{-0.02}^{+0.02}$ & \nodata & 560.3/508 \\
 3 & $1.45_{-0.08}^{+0.08}$ & $3.05_{-0.05}^{+0.06}$ & $6.3_{-0.2}^{+0.2}$ & \nodata & \nodata & \nodata & $1.42_{-0.03}^{+0.03}$ & $0.46_{-0.02}^{+0.02}$ & \nodata & 220.7/196 \\
 4 & $1.68_{-0.02}^{+0.02}$ & $2.703_{-0.012}^{+0.012}$ & $27.9_{-0.2}^{+0.2}$ & \nodata & \nodata & \nodata & $6.68_{-0.03}^{+0.03}$ & $1.866_{-0.018}^{+0.018}$ & \nodata & 817.7/700 \\
\cutinhead{Model: {\tt wabs(powerlaw + diskbb)}}
 1 & $1.41_{-0.13}^{+0.07}$ & $2.47_{-0.09}^{+0.05}$ & $20.2_{-2.4}^{+1.3}$ & \nodata & $0.292_{-0.019}^{+0.033}$ & $1.29_{-0.23}^{+0.19}$ & $6.67_{-0.06}^{+0.06}$ & $1.55_{-0.07}^{+0.07}$ & \nodata & 656.8/651 \\
 2 & $1.28_{-0.07}^{+0.09}$ & $2.23_{-0.05}^{+0.05}$ & $16.1_{-1.0}^{+1.0}$ & \nodata & $0.225_{-0.020}^{+0.019}$ & $2.4_{-0.4}^{+0.7}$ & $6.40_{-0.08}^{+0.08}$ & $1.38_{-0.05}^{+0.05}$ & \nodata & 518.0/506 \\
 3 & $1.39_{-0.12}^{+0.14}$ & $2.84_{-0.09}^{+0.09}$ & $5.2_{-0.5}^{+0.5}$ & \nodata & $0.17_{-0.02}^{+0.02}$ & $2.5_{-0.8}^{+1.3}$ & $1.45_{-0.03}^{+0.03}$ & $0.43_{-0.04}^{+0.04}$ & \nodata & 211.5/194 \\
 4 & $1.54_{-0.05}^{+0.04}$ & $2.55_{-0.03}^{+0.02}$ & $23.1_{-1.0}^{+0.7}$ & \nodata & $0.260_{-0.014}^{+0.018}$ & $1.48_{-0.19}^{+0.21}$ & $6.74_{-0.03}^{+0.03}$ & $1.71_{-0.04}^{+0.04}$ & \nodata & 756.4/698 \\
\cutinhead{Model: {\tt wabs$\ast$comptt}}
 1 & $0.67_{-0.09}^{+0.14}$ & $0.077_{-0.003}^{+0.003}$ & $0.66_{-0.06}^{+0.10}$ & [100] & $0.139_{-0.008}^{+0.006}$ & \nodata & $6.68_{-0.06}^{+0.05}$ & $1.12_{-0.04}^{+0.06}$ & 1.35 & 643.3/652 \\
 2 & $0.76_{-0.12}^{+0.08}$ & $0.107_{-0.004}^{+0.008}$ & $0.69_{-0.09}^{+0.08}$ & [100] & $0.116_{-0.008}^{+0.008}$ & \nodata & $6.28_{-0.05}^{+0.10}$ & $1.09_{-0.05}^{+0.03}$ & 1.41 & 524.1/507 \\
 3 & $0.79_{-0.18}^{+0.17}$ & $0.035_{-0.003}^{+0.004}$ & $0.28_{-0.06}^{+0.09}$ & [100] & $0.103_{-0.010}^{+0.010}$ & \nodata & $1.44_{-0.02}^{+0.03}$ & $0.29_{-0.03}^{+0.03}$ & 0.34 & 207.8/195 \\
 4 & $0.39_{-0.06}^{+0.06}$ & $4.3_{-0.4}^{+0.3}$ & $20._{-3.}^{+3.}$ & $2.9_{-0.3}^{+0.4}$ & $0.155_{-0.005}^{+0.005}$ & \nodata & $6.73_{-0.03}^{+0.03}$ & $1.035_{-0.019}^{+0.020}$ & 1.12 & 702.7/698 \\
\cutinhead{Model: {\tt wabs(comptt + diskbb)}}
 1 & $0.81_{-0.14}^{+0.10}$ & $0.076_{-0.003}^{+0.003}$ & $0.54_{-0.07}^{+0.11}$ & [100] & $0.159_{-0.021}^{+0.010}$ & $4.1_{-0.9}^{+0.7}$ & $6.66_{-0.06}^{+0.05}$ & $1.17_{-0.05}^{+0.05}$ & 1.53 & 641.2/651 \\
 2 & $0.96_{-0.08}^{+0.09}$ & $5.7_{-2.0}^{+1.0}$ & $13._{-6.}^{+3.}$ & $2.4_{-2.4}^{+1.1}$ & $0.195_{-0.015}^{+0.020}$ & $4.2_{-0.7}^{+1.0}$ & $6.31_{-0.10}^{+0.09}$ & $1.17_{-0.04}^{+0.05}$ & 1.51 & 514.5/505 \\
 3 & $0.97_{-0.13}^{+0.14}$ & $7.9_{-1.3}^{+3.2}$ & $9._{-2.}^{+2.}$ & $1.1_{-0.2}^{+0.4}$ & $0.165_{-0.018}^{+0.027}$ & $3.9_{-1.0}^{+1.3}$ & $1.41_{-0.03}^{+0.03}$ & $0.31_{-0.02}^{+0.03}$ & 0.46 & 203.2/193 \\
 4 & $0.75_{-0.06}^{+0.05}$ & $4.6_{-0.3}^{+0.5}$ & $15.7_{-2.0}^{+1.9}$ & $2.6_{-0.3}^{+0.4}$ & $0.192_{-0.010}^{+0.011}$ & $3.6_{-0.2}^{+0.2}$ & $6.72_{-0.03}^{+0.03}$ & $1.18_{-0.03}^{+0.02}$ & 1.43 & 677.6/697 \\
\enddata

\tablecomments{Columns are the same as Table~\ref{tab:ngc5204fit}.
}
\end{deluxetable*}

Holmberg II X-1 is surrounded by optical and radio nebulae \citep{pak02,mil05}. The optical nebula has a roughly isotropic morphology, and is thought to be photoionized requiring an input X-ray flux of at least $4 \times 10^{39}$~\ergs\ \citep{kaa04}. These indicate that it is a truly ULX with little beaming. The source has been observed four times with \xmm\ \citep[Table~\ref{tab:hoiiobs};][]{dew04,goa06}, and three times with \cha\ including one 100~ks observation (ObsID 5933) which has never been reported. Here, we re-analyzed all archival \xmm\ data, and included the 100~ks \cha\ data for timing analysis. We adopted a distance of 3.39~Mpc to the host galaxy \citep{kar02} and an absorption column density of $0.365 \times 10^{21}$~cm$^{-2}$ in our Galaxy.

\begin{figure}
\centering
\includegraphics[width=\columnwidth]{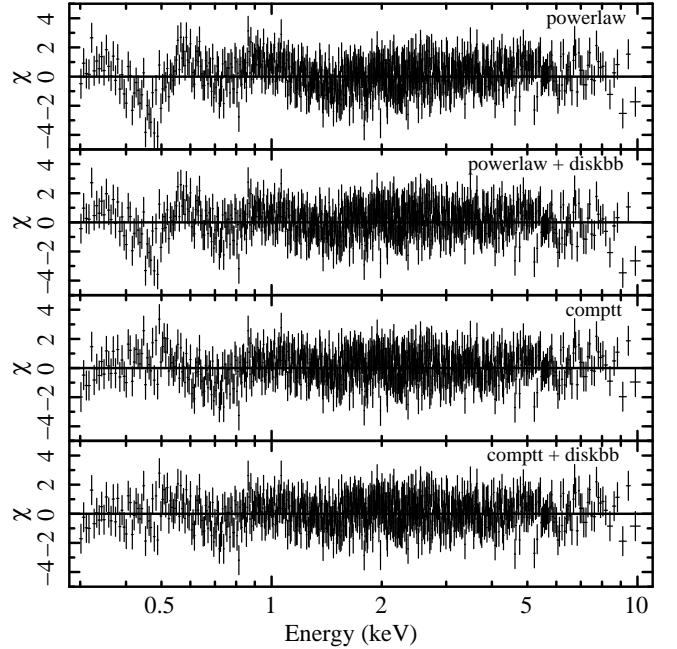}
\caption{
Data to model residuals for the 4th observation of Holmberg II X-1 in units of $\sigma$. Different panels are for different models with names shown on the upper right corner. Models containing a power-law component are unable to fit the data around 0.5~keV. Comptonization models provide good fits throughout the \xmm\ energy band.
\label{fig:hoiichi}}
\end{figure}

The radius of the circular region used to extract source spectra is 29\arcsec\ for the first two observations and 30\arcsec\ for the last two, respectively. The 4th observation has been reported in \citet{goa06}, in which the spectral analysis was made using a combination of 0.3--6~keV MOS data and 0.7--10~keV PN data in order to avoid calibration issues at other energy ranges. Here, we simply used 0.3--10 keV PN data for spectral fitting, since the statistics are good enough.

\begin{figure*}
\centering
\includegraphics[width=0.49\textwidth]{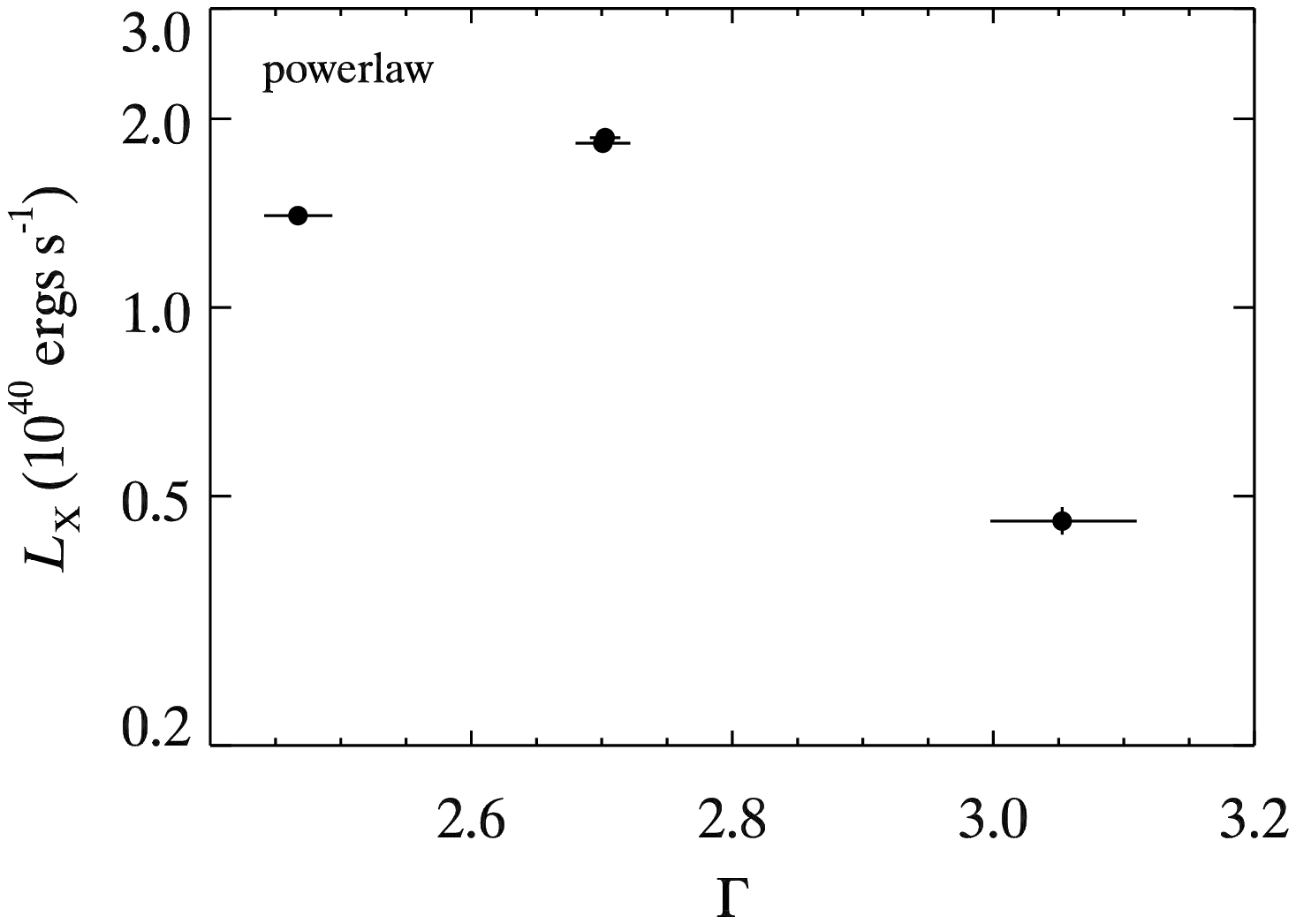}
\includegraphics[width=0.49\textwidth]{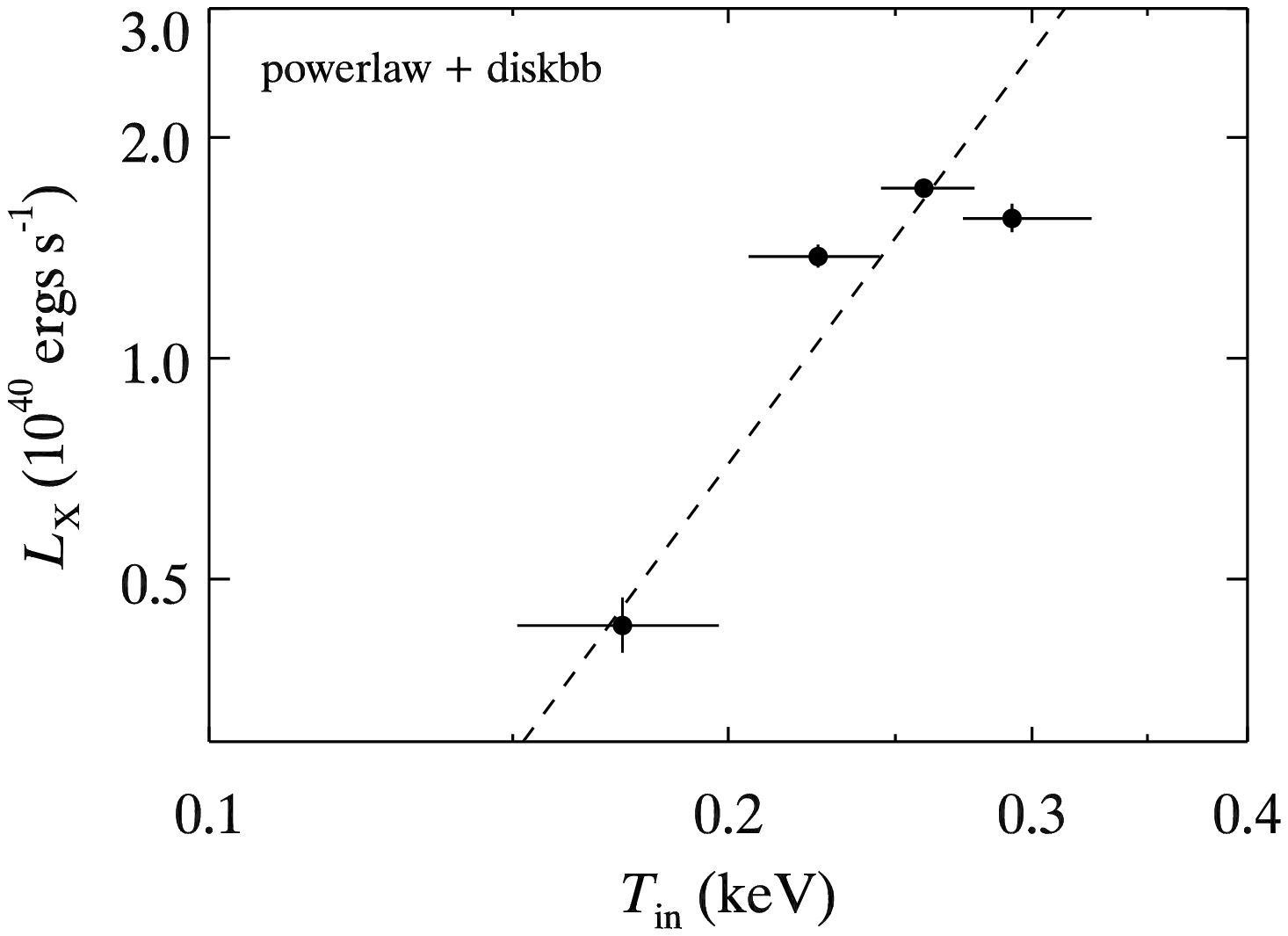}
\includegraphics[width=0.49\textwidth]{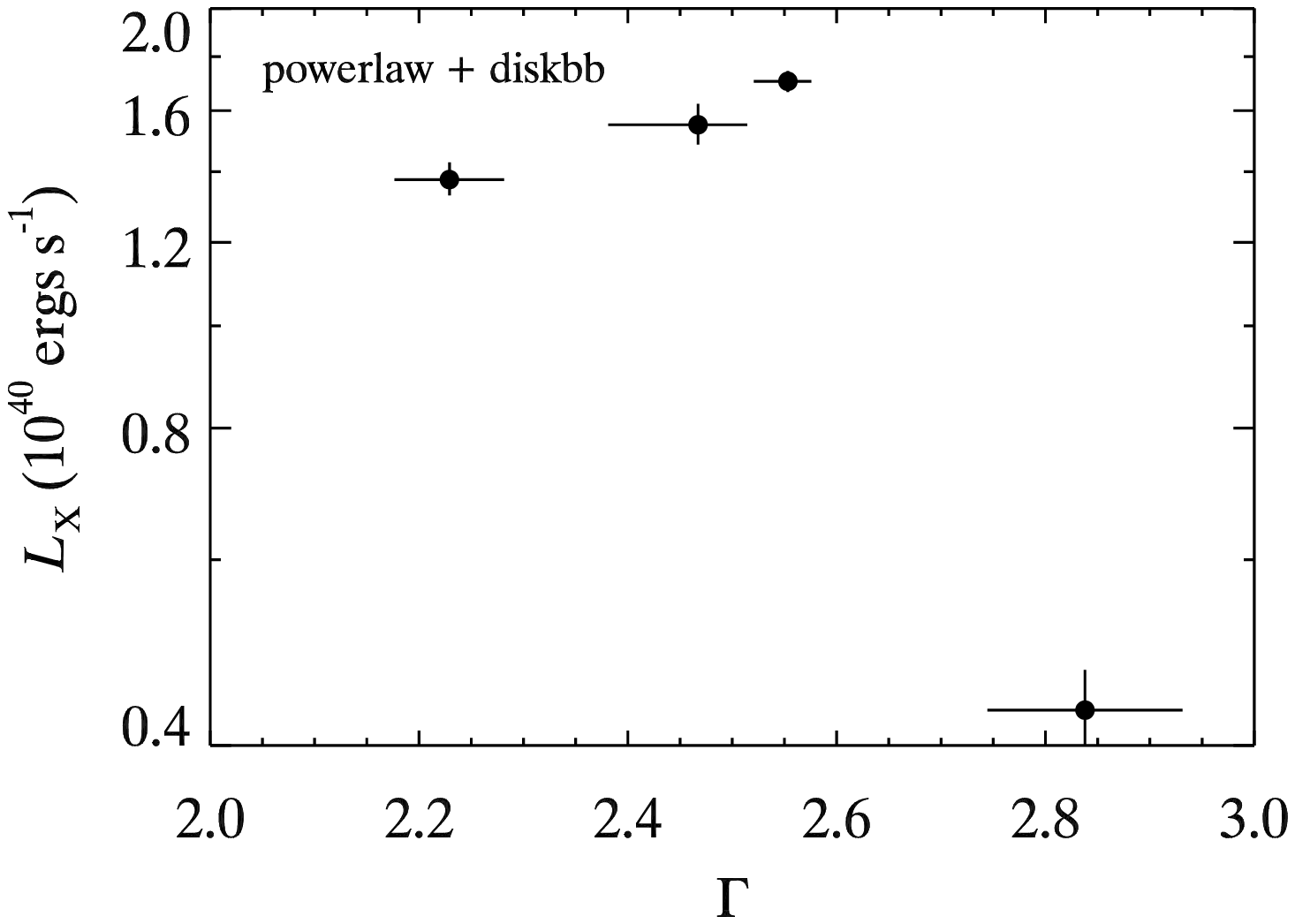}
\includegraphics[width=0.49\textwidth]{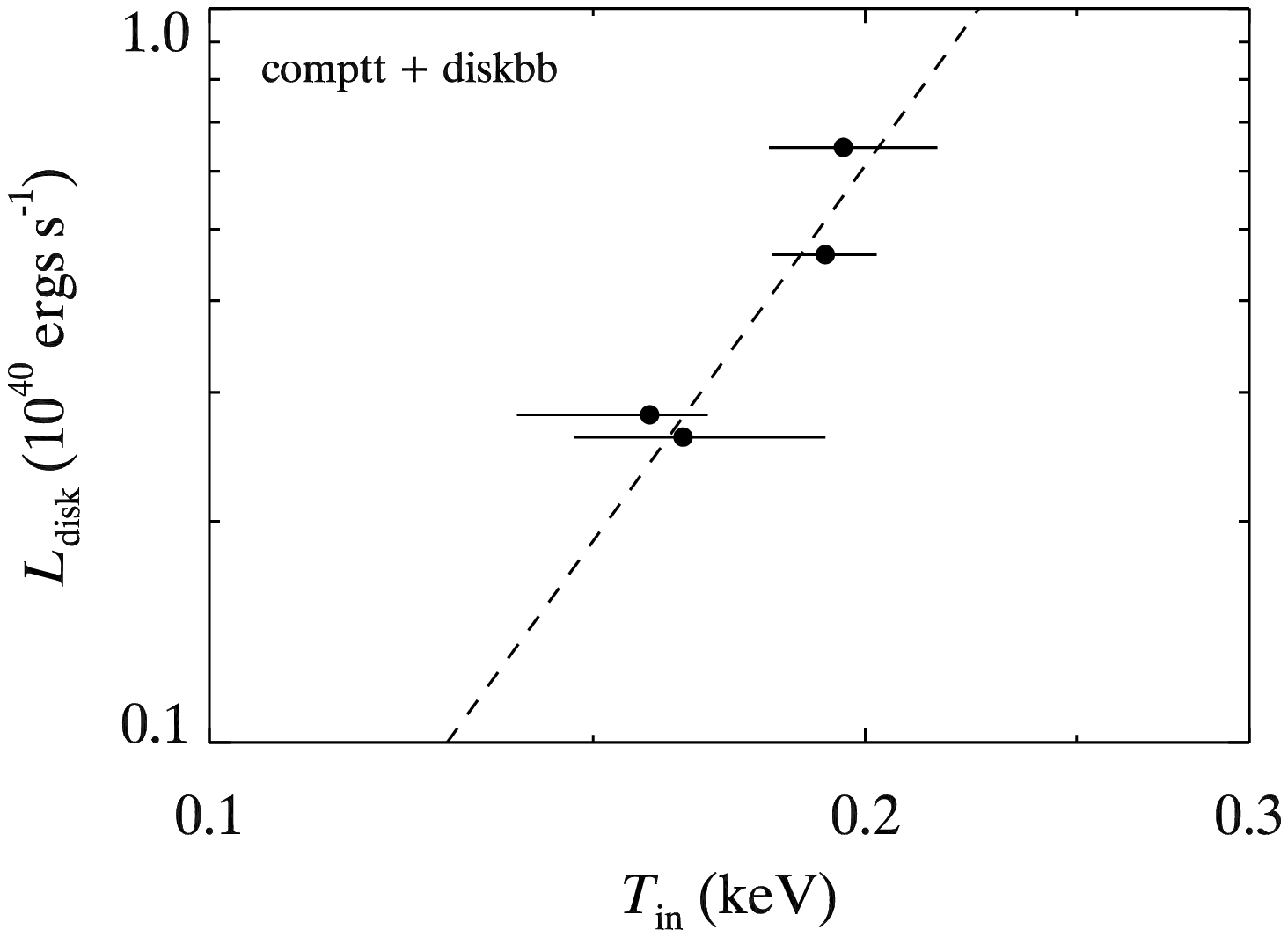}
\caption{
Diagrams between best-fit spectral parameters of Holmberg II X-1 derived from different models (data points adopted from Table~\ref{tab:hoiifit}). Dashed lines indicate best-fit relations: $L_{\rm X} \propto T_{\rm in}^{3.2 \pm 1.1}$, and $L_{\rm disk} \propto T_{\rm in}^{4.1 \pm 2.2}$. 
\label{fig:hoii}}
\end{figure*}

\begin{figure*}
\centering
\includegraphics[width=0.49\textwidth]{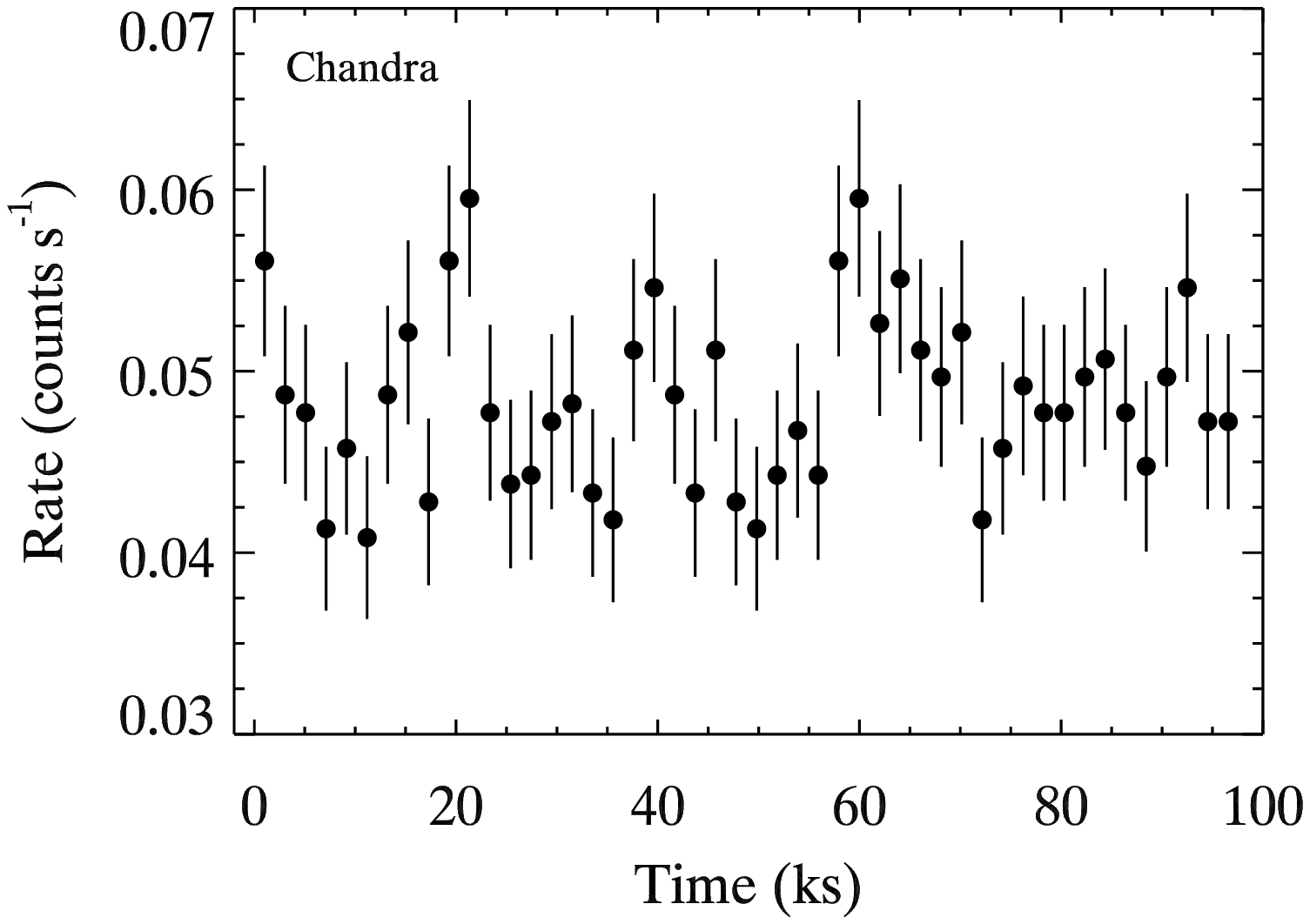}
\includegraphics[width=0.49\textwidth]{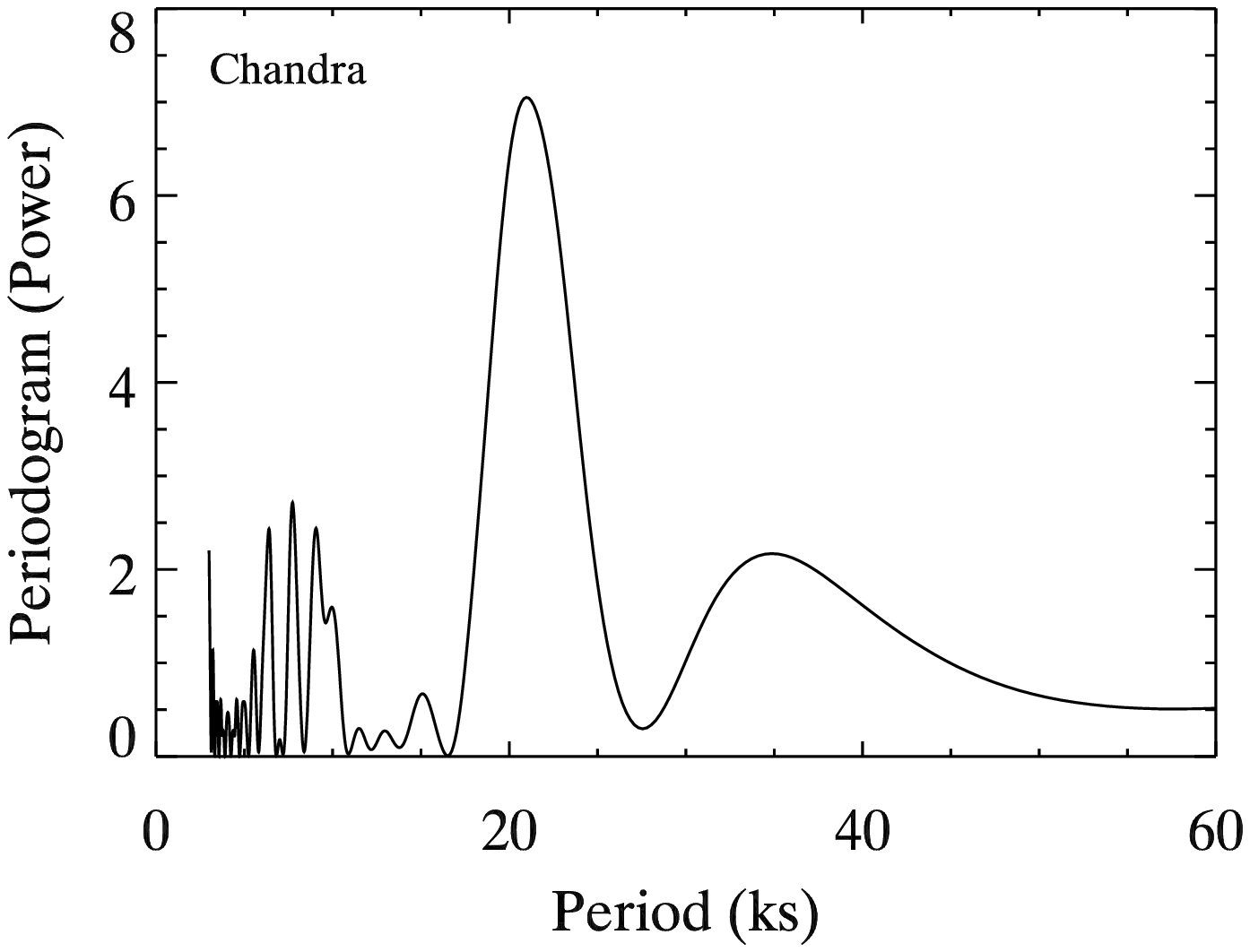}
\includegraphics[width=0.49\textwidth]{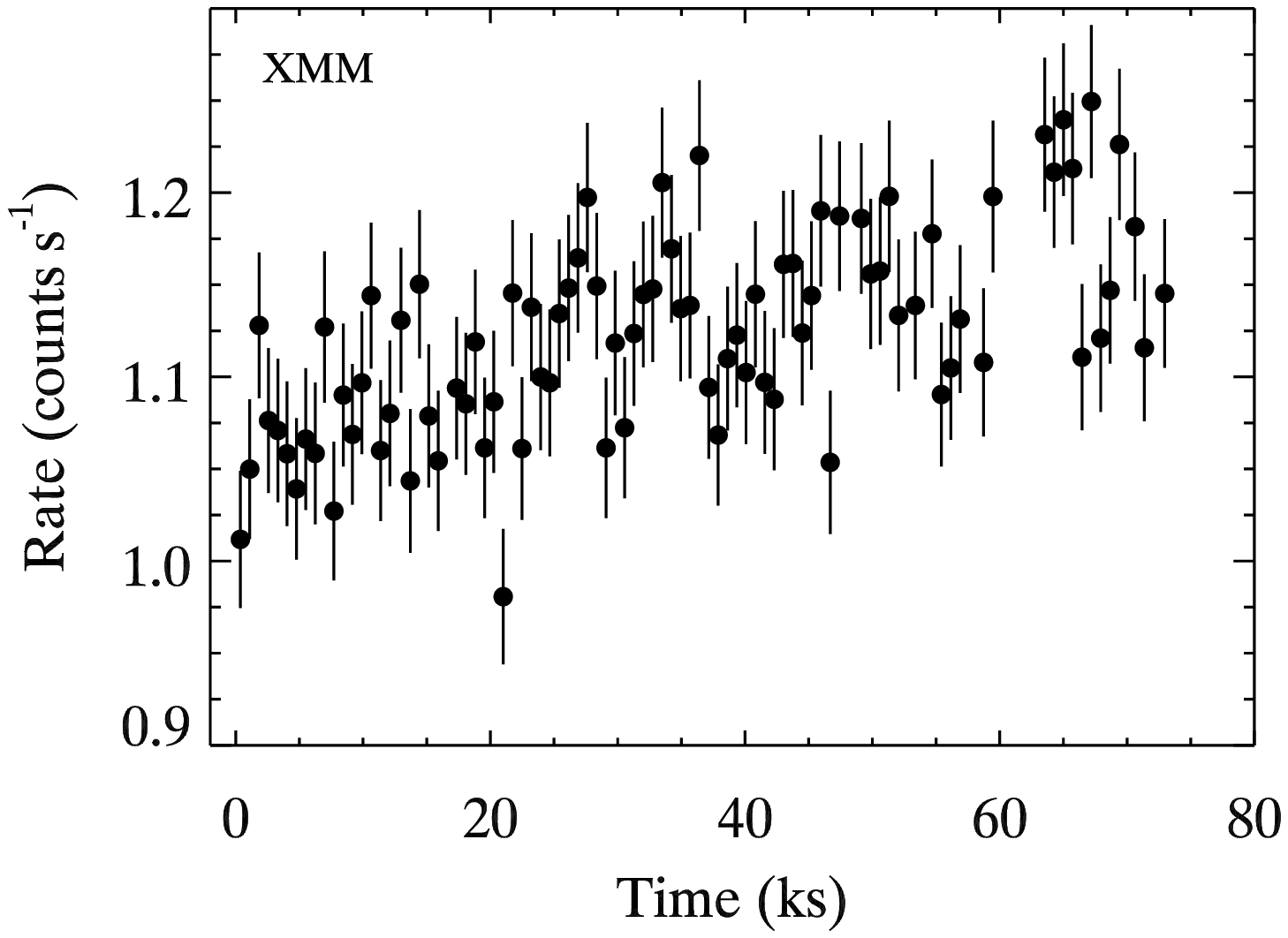}
\includegraphics[width=0.49\textwidth]{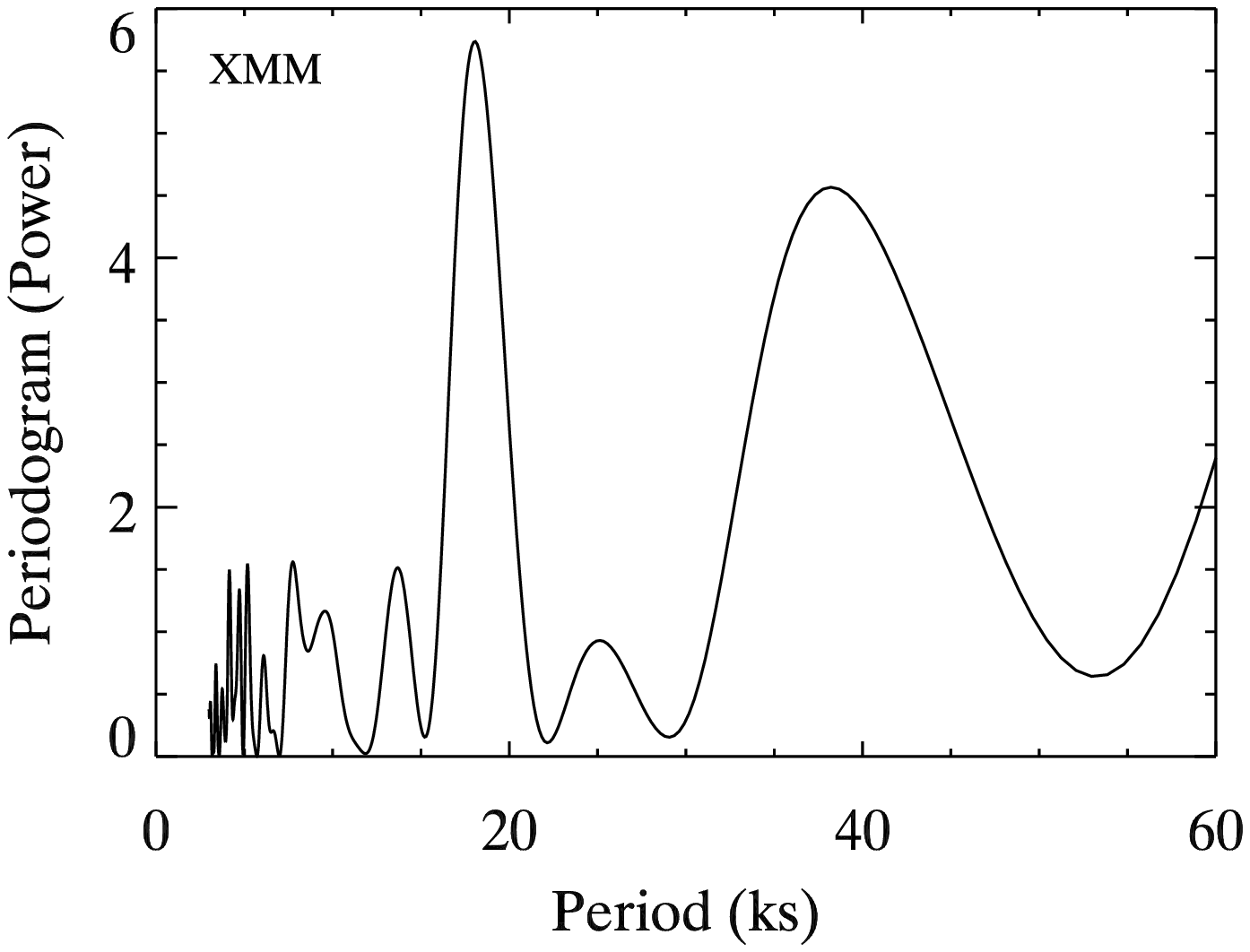}
\caption{
1--4.5 keV lightcurves and periodograms of Holmberg II X-1 obtained from \cha\ ACIS and \xmm\ PN observations. Timing gaps have been removed and background has been extracted for \xmm\ PN data.
\label{fig:hoiiv}}
\end{figure*}

\citet{goa06} suggested that the absorbing medium in the host galaxy has an oxygen abundance of 0.56 solar from spectral fits to RGS data. However, \citet{win07} found near-solar abundance of oxygen from EPIC data. As we have discussed previously, lower abundances systematically change the inferred luminosity and disk temperature but will not affect the confidence level of the correlation.  We therefore use solar abundance in the fits.

The same four models used for NGC 5204 X-1 were applied to Holmberg II X-1, and the best-fit parameters are listed in Table~\ref{tab:hoiifit}. The power-law model provides poor fits at energies around 0.5~keV, which could not be improved by adding a disk component (see Figure~\ref{fig:hoiichi}). In contrast, the Comptonization model, whether or not with a disk component, produces uniformly good fits in the full band. 

Diagrams between parameters of interest are plotted on Figure~\ref{fig:hoii}. Interestingly, the luminosity inferred from a disk plus power-law model goes higher when the disk temperature becomes hotter, with a relation of $L_{\rm X} \propto T_{\rm in}^{3.2 \pm 1.1}$, and a correlation coefficient of 0.84 with a chance probability of 0.16. This is consistent with the intrinsic evolution of disk blackbody emission, i.e., luminosity changes with the 4th power of the temperature. However, the four data points provide only weak evidence for the correlation. More observations are required to test it.

The Comptonization model provides better fits than the power-law model, especially at energies around 0.5~keV. For the first three observations, the electron temperature $T_{\rm e}$ is not well determined due to the lack of curvature in the spectrum at high energies and is thus fixed at 100~keV. No significant correlation can be seen between luminosity and seed temperature for a single Comptonization model. When we include a disk component, observation 2 and 3 also present possible high energy spectral curvature, and a plausible correlation between $L_{\rm disk}$ and $T_{\rm in}$. The best-fit relation is $L_{\rm disk} \propto T_{\rm in}^{4.1 \pm 2.2}$, and correlation coefficient between $\log L_{\rm disk}$ and $\log T_{\rm in}$ is 0.93 with a chance probability of 0.072. The best-fit inner radius of the disk is consistently around $4 \times 10^3$~km.

In Figure~\ref{fig:hoiiv}, we present lightcurves in the 1--4.5~keV energy range of Holmberg II X-1 from the longest \cha\ (ObsID 5933) and \xmm\ (no.\ 4) observation. For the \cha\ data, the lightcurve is extracted from a 3$\sigma$ elliptical region around the source found by {\tt wavdetect}, and binned at a timescale of 800 times the frame time (about 2.03~ks). No background is subtracted for the \cha\ data due to its small fraction. For the \xmm\ data, the lightcurve was extracted from the PN data only, corrected for background using the same region for spectral analysis, and binned at a timescale of 10,000 times the frame time (about 0.73~ks) at intervals without timing gaps. Strong variations are obviously seen from the lightcurves. Periodograms were calculated to examine typical timescales of the variation. A relatively narrow peak around 20~ks is pronounced, corresponding to multiple spikes in lightcurves at the same timescale. Due to the limited duration, we cannot determine whether or not the oscillation is a periodic signal or possibly a red noise fluctuation. We also examined variability at other energy ranges, and found they are not as prominent as in the 1--4.5~keV band.

\subsection{IC 342 X-1 and X-2}

\begin{deluxetable}{cccrrr}
\tablecolumns{6}
\tablewidth{0pc}
\tablecaption{\xmm\ observations of IC 342
\label{tab:ic342obs}}

\tablehead{
\colhead{} & \colhead{} & \colhead{} & \multicolumn{3}{c}{Good Exposures (ks)}\\
\cline{4-6}
\colhead{No.} & \colhead{ObsID} & \colhead{Date} & \colhead{PN} & \colhead{MOS1} & \colhead{MOS2}
}
\startdata
 1 & 0093640901 & 2001-02-11 & 4.8 & 9.5 & 9.5 \\
 2 & 0206890101 & 2004-02-20 & 10.3 & 20.4 & 21.4 \\
 3 & 0206890201 & 2004-08-17 & 17.1 & 23.4 & 23.4 \\
 4 & 0206890401 & 2005-02-10 & 4.6 & 7.3 & 7.6 \\
\enddata

\tablecomments{Good Exposures are effective exposures after background screening.}
\end{deluxetable}

\begin{deluxetable*}{ccccccccccc}
\tablecolumns{11}
\tablewidth{\textwidth}
\tablecaption{Best-fit parameters of IC 342 X-1
\label{tab:ic342x1fit}}

\tablehead{
\colhead{No.} & \colhead{$N_{\rm H}$} & \colhead{$\Gamma/\tau$} & \colhead{$N_{\rm PL}/N_{\rm C}$} & \colhead{$T_{\rm e}$} & \colhead{$T_{\rm in}$/$T_0$} & \colhead{$R_{\rm in}\sqrt{\cos i}$} & \colhead{$f_{\rm X}$} & \colhead{$L_{\rm X}$} & \colhead{$L_{\rm bol}$} & \colhead{$\chi^2$/dof}\\
\colhead{(1)} & \colhead{(2)} & \colhead{(3)} & \colhead{(4)} & \colhead{(5)} & \colhead{(6)} & \colhead{(7)} & \colhead{(8)} & \colhead{(9)} & \colhead{(10)} & \colhead{(11)}
}
\startdata
\multicolumn{11}{c}{Model: {\tt wabs$\ast$powerlaw}}\\ \noalign{\smallskip}\hline\noalign{\smallskip}
 1 & $5.8_{-0.3}^{+0.3}$ & $1.65_{-0.05}^{+0.05}$ & $5.0_{-0.3}^{+0.3}$ & \nodata & \nodata & \nodata & $2.51_{-0.07}^{+0.07}$ & $0.473_{-0.012}^{+0.012}$ & \nodata & 144.0/137 \\
 2 & $8.61_{-0.15}^{+0.16}$ & $2.04_{-0.02}^{+0.02}$ & $19.7_{-0.5}^{+0.5}$ & \nodata & \nodata & \nodata & $5.36_{-0.05}^{+0.05}$ & $1.41_{-0.02}^{+0.02}$ & \nodata & 750.2/667 \\
 3 & $6.32_{-0.15}^{+0.16}$ & $1.83_{-0.03}^{+0.03}$ & $6.9_{-0.2}^{+0.2}$ & \nodata & \nodata & \nodata & $2.64_{-0.04}^{+0.04}$ & $0.560_{-0.009}^{+0.009}$ & \nodata & 452.0/435 \\
 4 & $8.4_{-0.2}^{+0.2}$ & $1.89_{-0.03}^{+0.03}$ & $21.0_{-0.8}^{+0.8}$ & \nodata & \nodata & \nodata & $7.03_{-0.10}^{+0.10}$ & $1.64_{-0.03}^{+0.04}$ & \nodata & 421.0/374 \\
\cutinhead{Model: {\tt wabs(powerlaw + diskbb)}}
 1 & $6.9_{-1.0}^{+2.2}$ & $1.58_{-0.10}^{+0.11}$ & $4.6_{-0.7}^{+1.1}$ & \nodata & $0.28_{-0.09}^{+0.13}$ & $0.9_{-0.6}^{+3.3}$ & $2.54_{-0.08}^{+0.08}$ & $0.57_{-0.09}^{+0.28}$ & \nodata & 140.9/135 \\
 2 & $11.8_{-0.7}^{+1.0}$ & $2.17_{-0.03}^{+0.05}$ & $24.6_{-1.2}^{+1.9}$ & \nodata & $0.139_{-0.008}^{+0.007}$ & $24_{-5}^{+13}$ & $5.31_{-0.05}^{+0.05}$ & $4.1_{-1.0}^{+1.8}$ & \nodata & 734.4/665 \\
 3 & $7.5_{-0.5}^{+0.7}$ & $1.76_{-0.05}^{+0.05}$ & $6.3_{-0.5}^{+0.5}$ & \nodata & $0.26_{-0.04}^{+0.05}$ & $1.2_{-0.5}^{+1.0}$ & $2.67_{-0.04}^{+0.04}$ & $0.66_{-0.06}^{+0.10}$ & \nodata & 437.4/433 \\
 4 & $11.0_{-0.9}^{+1.4}$ & $2.02_{-0.04}^{+0.07}$ & $25.8_{-1.7}^{+3.0}$ & \nodata & $0.127_{-0.014}^{+0.010}$ & $31_{-10}^{+24}$ & $6.95_{-0.10}^{+0.10}$ & $4.6_{-1.4}^{+2.6}$ & \nodata & 409.8/372 \\
\cutinhead{Model: {\tt wabs$\ast$comptt}}
 1 & $4.7_{-0.8}^{+1.0}$ & $7.3_{-2.4}^{+1.1}$ & $3.7_{-1.6}^{+1.4}$ & $2.8_{-0.5}^{+3.0}$ & $0.23_{-0.07}^{+0.05}$ & \nodata & $2.47_{-0.09}^{+0.11}$ & $0.40_{-0.03}^{+0.04}$ & 0.51 & 140.6/135 \\
 2 & $7.9_{-2.1}^{+0.2}$ & $7.0_{-0.4}^{+1.1}$ & $24_{-11}^{+149}$ & $2.14_{-0.13}^{+0.17}$ & $0.14_{-0.14}^{+0.14}$ & \nodata & $5.21_{-0.06}^{+0.06}$ & $1.11_{-0.22}^{+0.12}$ & 1.26 & 711.2/665 \\
 3 & $5.2_{-0.4}^{+0.4}$ & $5.9_{-0.9}^{+0.5}$ & $4.0_{-1.0}^{+0.7}$ & $3.4_{-0.5}^{+1.2}$ & $0.22_{-0.03}^{+0.02}$ & \nodata & $2.64_{-0.04}^{+0.04}$ & $0.455_{-0.017}^{+0.019}$ & 0.58 & 439.8/433 \\
 4 & $7.9_{-0.3}^{+0.3}$ & $7.2_{-0.6}^{+0.6}$ & $50_{-23}^{+130}$ & $2.2_{-0.2}^{+0.3}$ & $0.06_{-0.06}^{+0.07}$ & \nodata & $6.80_{-0.11}^{+0.12}$ & $1.08_{-0.03}^{+0.05}$ & 1.83 & 409.8/372 \\
\enddata

\tablecomments{
Col.~(1): Observation index corresponding to Column (1) of Table~\ref{tab:ic342obs}.
Col.~(2): Absorption column density in units of $10^{21}$~cm$^{-2}$.
Col.~(3): $\Gamma$ is the photon index of the {\tt powerlaw} model; $\tau$ is the optical depth of the {\tt comptt} model.
Col.~(4): $N_{\rm PL}$ is the normalization of the {\tt powerlaw} model at 1~keV in units of $10^{-4}$~photons~cm$^{-2}$~s$^{-1}$; $N_{\rm C}$ is the normalization of the {\tt comptt} model in units of $10^{-4}$.
Col.~(5): Plasma temperature in units of keV; values in box brackets are fixed in the fitting.
Col.~(6): Inner disk temperature $T_{\rm in}$ of the {\tt diskbb} model or seed photon temperature $T_0$ of the {\tt comptt} model in units of keV.
Col.~(7): $R_{\rm in}$ is the inner disk radius in units of $10^3$~km; $i$ is the disk inclination angle.
Col.~(8): Absorbed flux in 0.3--10 keV in units of $10^{-12}$~\ergcms.
Col.~(9): Unabsorbed luminosity in 0.3--10 keV in units of $10^{40}$~\ergs.
Col.~(10): bolometric luminosity (integration of physical models in 0.01--100 keV) in units of $10^{40}$~\ergs.
Col.~(11): Best-fit $\chi^2$ and degrees of freedom.
All errors are at 1 $\sigma$ level.
}
\end{deluxetable*}

\begin{deluxetable*}{cccccccccc}
\tablecolumns{10}
\tablewidth{\textwidth}
\tablecaption{Best-fit parameters of IC 342 X-2. The spectral fits were performed in the 1--10 keV range for the {\tt wabs$\ast$powerlaw} and {\tt wabs$\ast$diskbb} models, but in the 0.5--10 keV range for the {\tt wabs(powerlaw + apec)} model.
\label{tab:ic342x2fit}}

\tablehead{
\colhead{No.} & \colhead{$N_{\rm H}$} & \colhead{$\Gamma$} & \colhead{$N_{\rm PL}$} & \colhead{$T_{\rm in}$/$kT$} & \colhead{$R_{\rm in}\sqrt{\cos i}$/$N_{\rm A}$} & \colhead{$f_{\rm X}$} & \colhead{$L_{\rm X}$} & \colhead{$L_{\rm bol}$} & \colhead{$\chi^2$/dof}\\
\colhead{(1)} & \colhead{(2)} & \colhead{(3)} & \colhead{(4)} & \colhead{(5)} & \colhead{(6)} & \colhead{(7)} & \colhead{(8)} & \colhead{(9)} & \colhead{(10)}
}
\startdata
\multicolumn{10}{c}{Model: {\tt wabs$\ast$powerlaw}}\\ \noalign{\smallskip}\hline\noalign{\smallskip}
 1 & $2.3_{-0.2}^{+0.3}$ & $1.79_{-0.13}^{+0.14}$ & $8.3_{-1.6}^{+2.0}$ & \nodata & \nodata & $2.58_{-0.11}^{+0.12}$ & $0.51_{-0.03}^{+0.03}$ & \nodata & 45.0/39 \\
 2 & $2.53_{-0.06}^{+0.07}$ & $1.71_{-0.03}^{+0.03}$ & $30.2_{-1.4}^{+1.7}$ & \nodata & \nodata & $10.55_{-0.11}^{+0.10}$ & $2.08_{-0.03}^{+0.03}$ & \nodata & 717.2/605 \\
 3 & $1.79_{-0.04}^{+0.08}$ & $1.37_{-0.04}^{+0.04}$ & $3.9_{-0.2}^{+0.3}$ & \nodata & \nodata & $2.50_{-0.04}^{+0.04}$ & $0.422_{-0.007}^{+0.007}$ & \nodata & 350.6/316 \\
 4 & $2.14_{-0.16}^{+0.21}$ & $1.64_{-0.08}^{+0.11}$ & $6.4_{-0.8}^{+1.2}$ & \nodata & \nodata & $2.59_{-0.08}^{+0.08}$ & $0.485_{-0.018}^{+0.020}$ & \nodata & 74.5/84 \\
\cutinhead{Model: {\tt wabs$\ast$diskbb}}
 1 & $1.6_{-0.2}^{+0.2}$ & \nodata & \nodata & $2.17_{-0.18}^{+0.22}$ & $3.0_{-0.5}^{+0.6}$ & $2.41_{-0.11}^{+0.12}$ & $0.408_{-0.018}^{+0.019}$ & 0.50 & 34.9/37 \\
 2 & $1.82_{-0.04}^{+0.04}$ & \nodata & \nodata & $2.50_{-0.05}^{+0.06}$ & $4.64_{-0.19}^{+0.19}$ & $10.14_{-0.11}^{+0.11}$ & $1.707_{-0.017}^{+0.017}$ & 2.19 & 591.7/596 \\
 3 & $1.35_{-0.04}^{+0.08}$ & \nodata & \nodata & $3.34_{-0.19}^{+0.11}$ & $1.31_{-0.07}^{+0.13}$ & $2.43_{-0.04}^{+0.04}$ & $0.378_{-0.006}^{+0.006}$ & 0.55 & 303.6/311 \\
 4 & $1.42_{-0.15}^{+0.12}$ & \nodata & \nodata & $2.61_{-0.15}^{+0.27}$ & $2.1_{-0.3}^{+0.2}$ & $2.49_{-0.09}^{+0.09}$ & $0.397_{-0.012}^{+0.012}$ & 0.52 & 70.8/83 \\
\cutinhead{Model: {\tt wabs(powerlaw + apec)}}
 2 & $3.22_{-0.10}^{+0.10}$ & $1.91_{-0.04}^{+0.04}$ & $44_{-3}^{+3}$ & $0.092_{-0.003}^{+0.008}$ & $51_{-27}^{+33}$ & $10.41_{-0.11}^{+0.11}$ & $3.6_{-0.3}^{+0.4}$ & \nodata & 603.2/608 \\
 3 & $2.58_{-0.15}^{+0.15}$ & $1.59_{-0.06}^{+0.05}$ & $5.9_{-0.5}^{+0.6}$ & $0.126_{-0.012}^{+0.016}$ & $0.4_{-0.3}^{+0.6}$ & $2.47_{-0.04}^{+0.04}$ & $0.60_{-0.05}^{+0.06}$ & \nodata & 327.6/320 \\
\enddata

\tablecomments{
Col.~(1): Observation index corresponding to Column (1) of Table~\ref{tab:ic342obs}.
Col.~(2): Absorption column density in units of $10^{22}$~cm$^{-2}$.
Col.~(3): Photon index of the {\tt powerlaw} model.
Col.~(4): Normalization of the {\tt powerlaw} model at 1~keV in units of $10^{-4}$~photons~cm$^{-2}$~s$^{-1}$.
Col.~(5): $T_{\rm in}$ is the inner disk temperature  of the {\tt diskbb} model; $kT$ is the plasma temperature of the {\tt apec} model. Both are in units of keV. 
Col.~(6): $R_{\rm in}$ is the inner disk radius in units of $10^3$~km and $i$ is the disk inclination angle for the {\tt diskbb} model; $N_{\rm A}$ is the normalization of the {\tt apec} model. 
Col.~(7): Absorbed flux in 1--10 keV in units of $10^{-12}$~\ergcms.
Col.~(8): Unabsorbed luminosity in 1--10 keV in units of $10^{40}$~\ergs.
Col.~(9): bolometric luminosity (integration of physical models in 0.01--100 keV) in units of $10^{40}$~\ergs.
Col.~(10): Best-fit $\chi^2$ and degrees of freedom.
All errors are at 1 $\sigma$ level.
}
\end{deluxetable*}

\begin{figure}
\centering
\includegraphics[width=\columnwidth]{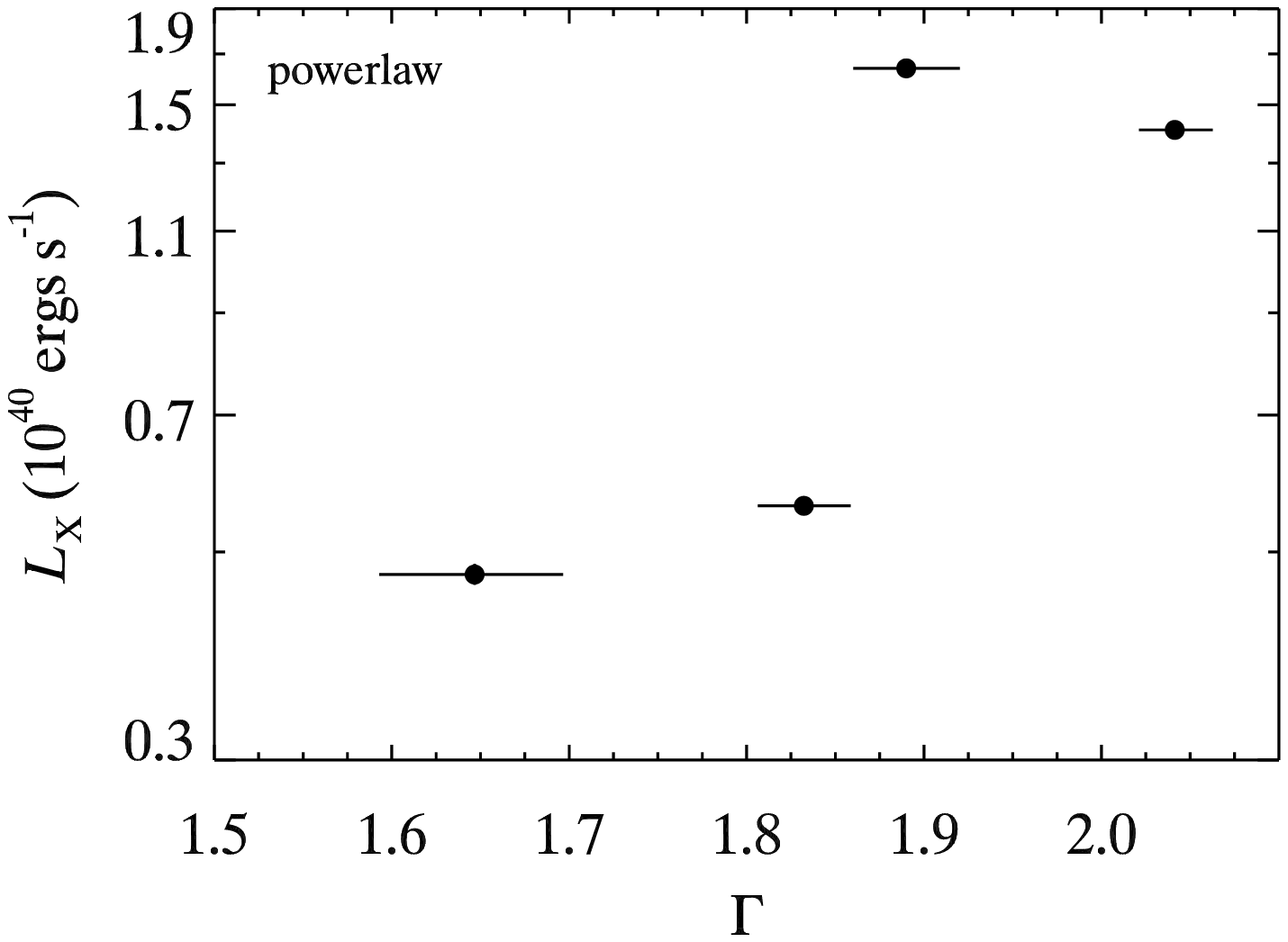}
\includegraphics[width=\columnwidth]{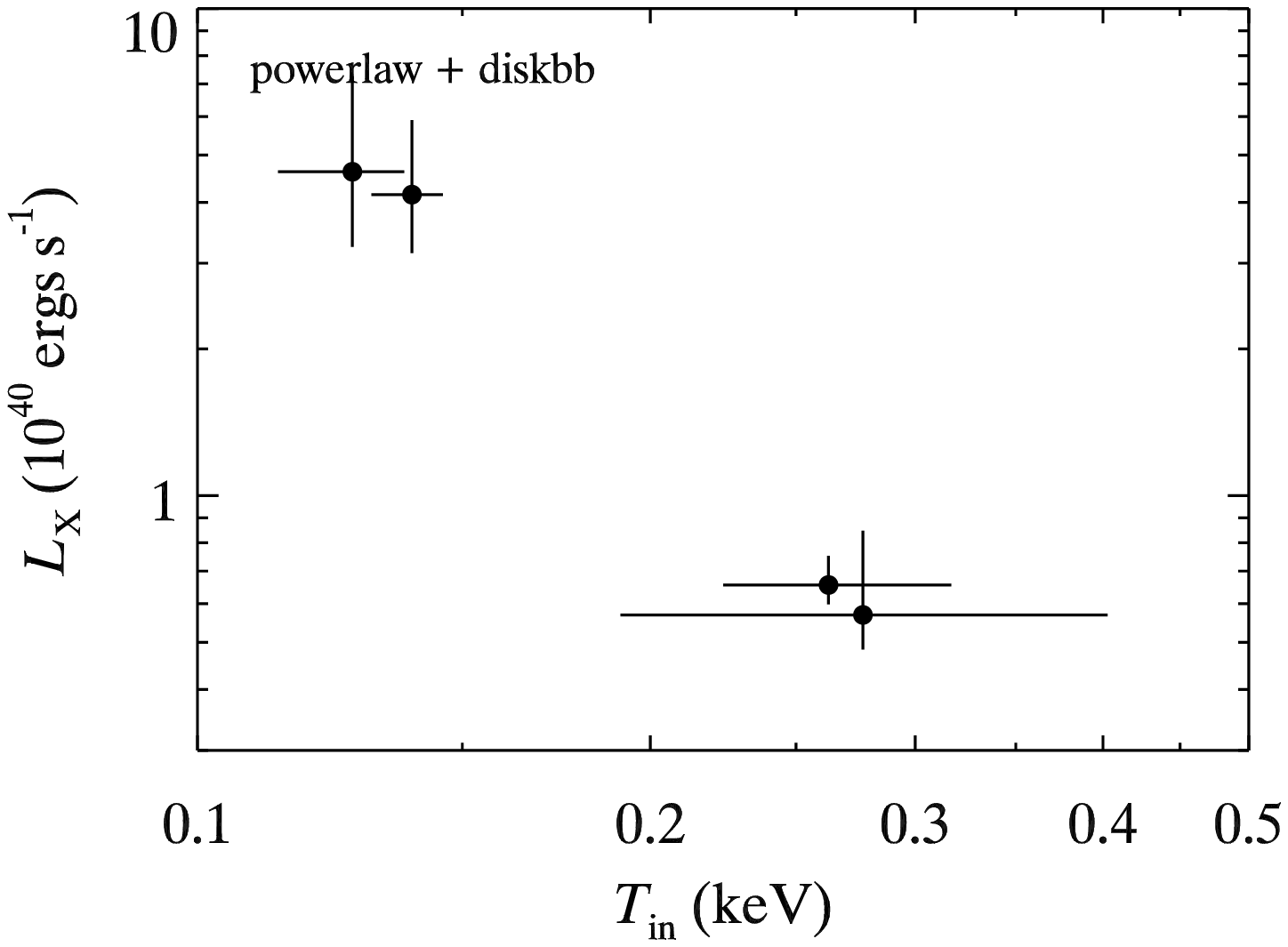}
\caption{
$L_{\rm X}$ versus $T_{\rm in}$ of IC 342 X-1 derived from the disk plus power-law model (data points adopted from Table~\ref{tab:ic342x1fit}). The four data points represent a relation of $L_{\rm X} \propto T_{\rm in}^{-2.8 \pm 0.7}$.
\label{fig:ic342x1}}
\end{figure}

\begin{figure}
\centering
\includegraphics[width=\columnwidth]{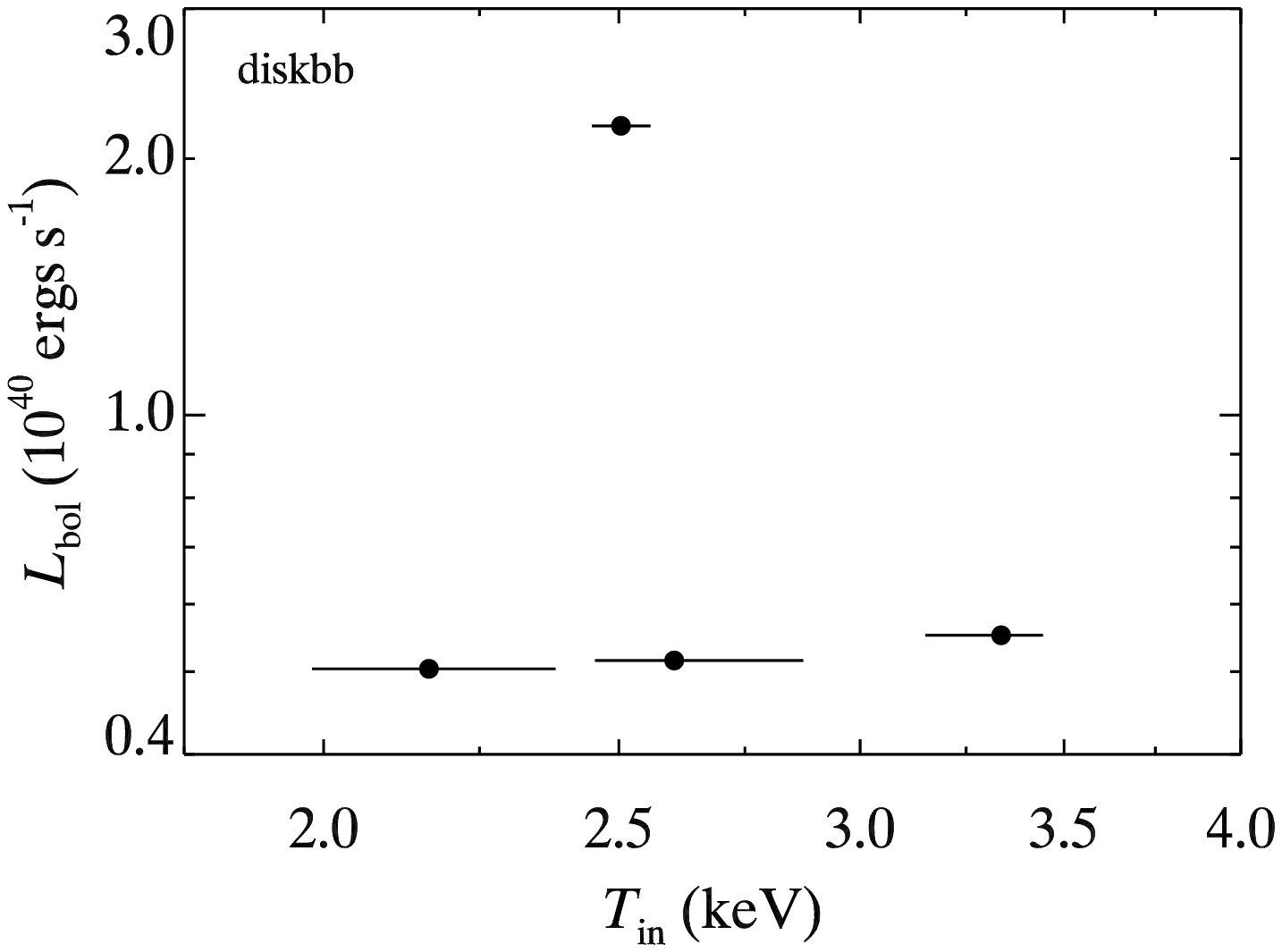}
\caption{
$L_{\rm bol}$ versus $T_{\rm in}$ of IC 342 X-2 derived from the absorbed disk model (data points adopted from Table~\ref{tab:ic342x2fit}). The spectral fits were performed in the energy range of 1--10 keV to avoid the soft excess whose origin is unclear.
\label{fig:ic342x2}}
\end{figure}

\begin{figure}
\centering
\includegraphics[width=0.8\columnwidth]{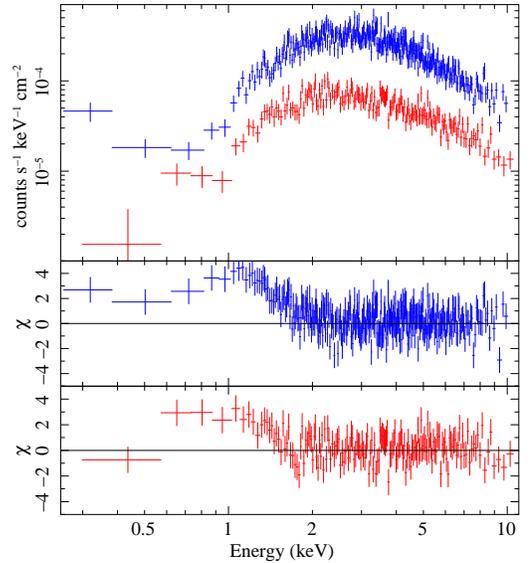}
\caption{
\xmm\ PN spectra of IC 342 X-2 in observations 2 (blue) and 3 (red). Measured spectra are shown in the {\it top} panel, and their residuals (in units of $\sigma$) after fitting with a power-law model in 2--10 keV are shown in the {\it middle} and {\it bottom} panels. An soft excess between 0.5--1.5 keV is obvious in both spectra, which can be adequately fitted by the thermal plasma model {\tt apec}. The high state spectrum, however, present flux excess below 0.5~keV, which seems to be a super soft emission component and not subject to the interstellar absorption. 
\label{fig:ic342x2hl}}
\end{figure}

IC 342 X-1 and X-2 are famous because they are the first two ULXs in which state transitions were discovered \citep{kub01}, and are thus ideal objects to study here. IC 342 X-1 resides in an optical nebula \citep{pak02,rob03,fen08}, which is similar to Holmberg II X-1. The galaxy has been observed four times with \xmm\ (Table~\ref{tab:ic342obs}), and the last three observations have never been reported.

The radius used to extract spectra for X-1 is 25\arcsec\ for the 1st observation and 30\arcsec\ for the other three; for X-2, it is 20\arcsec\ for the 2nd observation and 32\arcsec\ for the rest. For X-2 in observation 1, the MOS data are not used because the source is located right across the CCD gap. We adopt a distance of 3.3~Mpc to the host galaxy \citep{sah02}, and an absorption column density of $3.08 \times 10^{21}$~cm$^{-2}$ in our Galaxy.

We tried three different models to fit the spectra of X-1; the disk plus Comptonization model is not used because a global minimum cannot be found for this model in some observations due to the weakness of the disk component. The best-fit parameters are listed in Table~\ref{tab:ic342x1fit}. If we fit the data with an absorbed power-law model, the four observations appear in two groups: a low hard group and a high soft group (see Figure~\ref{fig:ic342x1}). The absorption column density is also correlated with the source luminosity. When the disk component is added, the disk temperature seems to be inversely scaled with the luminosity with a best-fit relation of $L_{\rm X} \propto T_{\rm in}^{-2.8 \pm 0.7}$, which excludes a $L \propto T^4$ relation. 
Fitting with the Comptonization model, a cool, optically thick corona is derived as a consequence of spectral curvature at energies close to 10~keV. This is especially evident in the 2nd observation, in which the Comptonization model significantly improved the fits relative to the disk plus power-law model. However, no correlation can be found between spectral parameters with the Comptonization model due to the relatively large errors.

The shape of the energy spectrum of X-2 seems to be abnormal. We plotted the source spectra in the 2nd and 3rd observations in Figure~\ref{fig:ic342x2hl}, as well as residuals after fitting with an absorbed power-law model to the data in 2--10 keV. Excesses at low energies are seen in the residuals. Unlike usual soft excesses found in ULXs \citep[e.g., Fig.~2 in][]{gon06}, these excesses peak around 1~keV and have a minimum around 0.5~keV; they cannot be fitted by a cool disk component but can be explained by thermal plasma emission ({\tt apec} in XSPEC) down to 0.5~keV. Therefore, we first tried a power-law model and an accretion disk model modified by absorption to fit the data in the energy range of 1--10 keV to avoid the soft excesses. Since an {\tt apec} model in addition to the power-law is able to fit the data down to 0.5~keV, we then applied such a model to the 2nd and 3rd observations in the 0.5--10~keV range; the 1st and 4th observations do not have enough counts to allow significant detection of the {\tt apec} component. We note that we cannot find a physical model that is able to adequately fit the data down to 0.3~keV for the 2nd observation. Below 0.5~keV, there seems to be another emission component that is not subject to the absorption. All best-fit spectral parameters for different models are listed in Table~\ref{tab:ic342x2fit}. To be consistent, $f_{\rm X}$ and $L_{\rm X}$ are all quoted in the energy range of 1--10 keV.

Two distinct flux states are found in the four observations, but no correlation is shown between spectral parameters. Unlike that reported in \citet{kub01}, the low state and the high state has almost the same spectral shape except at energies below 0.7~keV (Figure~\ref{fig:ic342x2hl}). Fitting with a single accretion disk model indicates that the source stays in the thermal dominant state, in which the $L \propto T^4$ relation should robustly exist. However, the derived $L_{\rm bol}$ and $T_{\rm in}$ strongly violate the $L \propto T^4$ relation (see Figure~\ref{fig:ic342x2}); fitting with a $L \propto T^4$ function results in $\chi^2=120.9$ with 3 degrees of freedom. This is solid evidence against the hot disk model, indicating the source emission is not from a standard, hot accretion disk onto a stellar-mass black hole. The {\tt apec} model is able to explain the soft excess between 0.5--1.4~keV. In the 1--10 keV range, the {\tt apec} component contributes 38\% and 22\% of the total luminosity, respectively for the 2nd and 3rd observation.

\subsection{Antennae X-11, X-16, X-42, and X-44}

\begin{deluxetable}{cccrrr}
\tablecolumns{6}
\tablewidth{0pc}
\tablecaption{\xmm\ observations of the Antennae
\label{tab:antobs}}

\tablehead{
\colhead{} & \colhead{} & \colhead{} & \multicolumn{3}{c}{Good Exposures (ks)}\\
\cline{4-6}
\colhead{No.} & \colhead{ObsID} & \colhead{Date} & \colhead{PN} & \colhead{MOS1} & \colhead{MOS2}
}
\startdata
 1 & 0085220101 & 2002-01-18 & 14.9 & 19.9 & 20.0 \\
 2 & 0085220201 & 2002-01-08 & 33.6 & 51.2 & 51.4 \\
 3 & 0500070201 & 2007-06-09 & 9.1 & 13.5 & 13.6 \\
 4 & 0500070301 & 2007-06-24 & 19.2 & 24.1 & 23.9 \\
 5 & 0500070401 & 2007-12-10 & 8.4 & 18.4 & 19.7 \\
 6 & 0500070501 & 2007-07-09 & 14.6 & 24.2 & 10.3 \\
 7 & 0500070601 & 2007-12-18 & 17.4 & 25.1 & 24.4 \\
 8 & 0500070701 & 2007-12-26 & 21.7 & 31.1 & 31.0 \\
\enddata

\tablecomments{Good Exposures are effective exposures after background screening.}
\end{deluxetable}

\begin{deluxetable*}{ccccccc}[bht]
\tablecolumns{7}
\tablewidth{\textwidth}
\tablecaption{Best-fit parameters of Antennae X-11/16/42/44 with an absorbed power-law model, {\tt wabs(powerlaw + apec)}, where {\tt apec} parameters are fixed to serve as a background spectrum.
\label{tab:antfit}}

\tablehead{
\colhead{No.} & \colhead{$N_{\rm H}$} & \colhead{$\Gamma$} & \colhead{$N_{\rm PL}$} & \colhead{$f_{\rm X}$} & \colhead{$L_{\rm X}$} & \colhead{$\chi^2$/dof}\\
\colhead{(1)} & \colhead{(2)} & \colhead{(3)} & \colhead{(4)} & \colhead{(5)} & \colhead{(6)} & \colhead{(7)}
}
\startdata
\multicolumn{7}{c}{X-11}\\ \noalign{\smallskip}\hline\noalign{\smallskip}
 1 & $0.31^{<0.51}$ & $1.61_{-0.09}^{+0.12}$ & $0.305_{-0.019}^{+0.042}$ & $0.218_{-0.020}^{+0.014}$ & $1.33_{-0.09}^{+0.10}$ & 20.4/17 \\
 2 & $1.16_{-0.11}^{+0.12}$ & $1.76_{-0.05}^{+0.05}$ & $0.55_{-0.04}^{+0.04}$ & $0.301_{-0.011}^{+0.011}$ & $2.11_{-0.07}^{+0.07}$ & 67.1/67 \\
 3 & $1.24_{-0.17}^{+0.17}$ & $1.56_{-0.08}^{+0.08}$ & $0.39_{-0.04}^{+0.04}$ & $0.262_{-0.017}^{+0.017}$ & $1.76_{-0.09}^{+0.09}$ & 23.7/29 \\
 4 & $0.75_{-0.11}^{+0.10}$ & $1.68_{-0.06}^{+0.06}$ & $0.34_{-0.02}^{+0.02}$ & $0.209_{-0.010}^{+0.010}$ & $1.38_{-0.05}^{+0.05}$ & 56.2/59 \\
 5 & $0.40^{<1.95}$ & $1.3_{-0.3}^{+0.3}$ & $0.13_{-0.03}^{+0.08}$ & $0.13_{-0.03}^{+0.02}$ & $0.79_{-0.10}^{+0.11}$ & 2.4/6 \\
 6 & $1.24_{-0.19}^{+0.16}$ & $1.79_{-0.13}^{+0.13}$ & $0.28_{-0.03}^{+0.04}$ & $0.147_{-0.012}^{+0.013}$ & $1.05_{-0.06}^{+0.06}$ & 39.6/29 \\
 7 & $0.32^{<0.66}$ & $1.40_{-0.13}^{+0.18}$ & $0.150_{-0.016}^{+0.029}$ & $0.134_{-0.014}^{+0.013}$ & $0.81_{-0.07}^{+0.07}$ & 19.1/12 \\
 8 & $0.33^{<0.56}$ & $1.83_{-0.12}^{+0.14}$ & $0.183_{-0.017}^{+0.036}$ & $0.105_{-0.009}^{+0.010}$ & $0.66_{-0.05}^{+0.05}$ & 29.8/16 \\
\cutinhead{X-16}
 1 & $0.47_{-0.11}^{+0.12}$ & $1.26_{-0.07}^{+0.06}$ & $0.27_{-0.02}^{+0.02}$ & $0.284_{-0.015}^{+0.015}$ & $1.72_{-0.08}^{+0.08}$ & 31.9/44 \\
 2 & $0.46_{-0.07}^{+0.07}$ & $1.35_{-0.04}^{+0.03}$ & $0.318_{-0.015}^{+0.015}$ & $0.298_{-0.009}^{+0.009}$ & $1.82_{-0.05}^{+0.05}$ & 120.4/119 \\
 3 & $0.31^{<0.38}$ & $1.20_{-0.10}^{+0.14}$ & $0.133_{-0.010}^{+0.014}$ & $0.159_{-0.010}^{+0.015}$ & $0.90_{-0.08}^{+0.10}$ & 13.8/12 \\
 4 & $0.31^{<0.39}$ & $1.22_{-0.09}^{+0.09}$ & $0.098_{-0.007}^{+0.008}$ & $0.112_{-0.010}^{+0.008}$ & $0.68_{-0.04}^{+0.04}$ & 14.4/18 \\
 5 & $0.31^{<0.50}$ & $1.11_{-0.05}^{+0.08}$ & $0.218_{-0.011}^{+0.022}$ & $0.283_{-0.017}^{+0.013}$ & $1.68_{-0.09}^{+0.09}$ & 24.4/29 \\
 6 & $0.86_{-0.14}^{+0.15}$ & $1.31_{-0.08}^{+0.07}$ & $0.22_{-0.02}^{+0.02}$ & $0.215_{-0.012}^{+0.013}$ & $1.34_{-0.07}^{+0.07}$ & 16.6/30 \\
 7 & $0.78_{-0.11}^{+0.11}$ & $1.29_{-0.05}^{+0.05}$ & $0.30_{-0.02}^{+0.02}$ & $0.301_{-0.014}^{+0.014}$ & $1.87_{-0.07}^{+0.07}$ & 67.2/57 \\
 8 & $0.37^{<0.44}$ & $1.20_{-0.05}^{+0.05}$ & $0.242_{-0.013}^{+0.015}$ & $0.280_{-0.011}^{+0.011}$ & $1.68_{-0.06}^{+0.06}$ & 73.0/68 \\
\cutinhead{X-42}
 1 & $0.80_{-0.14}^{+0.14}$ & $1.73_{-0.11}^{+0.10}$ & $0.25_{-0.03}^{+0.03}$ & $0.144_{-0.011}^{+0.012}$ & $0.96_{-0.05}^{+0.05}$ & 31.0/31 \\
 2 & $0.63_{-0.08}^{+0.08}$ & $1.66_{-0.06}^{+0.05}$ & $0.241_{-0.016}^{+0.017}$ & $0.155_{-0.006}^{+0.006}$ & $1.00_{-0.03}^{+0.03}$ & 83.4/86 \\
 3 & $1.4_{-0.7}^{+0.3}$ & $2.2_{-0.5}^{+0.3}$ & $0.23_{-0.09}^{+0.06}$ & $0.083_{-0.011}^{+0.014}$ & $0.65_{-0.07}^{+0.16}$ & 14.7/12 \\
 4 & $1.16_{-0.21}^{+0.18}$ & $2.21_{-0.11}^{+0.09}$ & $0.23_{-0.03}^{+0.03}$ & $0.081_{-0.005}^{+0.005}$ & $0.67_{-0.06}^{+0.07}$ & 46.6/31 \\
 5 & $0.65_{-0.15}^{+0.17}$ & $1.70_{-0.10}^{+0.10}$ & $0.27_{-0.03}^{+0.04}$ & $0.168_{-0.011}^{+0.012}$ & $1.10_{-0.06}^{+0.06}$ & 35.3/26 \\
 6 & $0.53_{-0.17}^{+0.16}$ & $1.75_{-0.17}^{+0.18}$ & $0.19_{-0.03}^{+0.03}$ & $0.117_{-0.010}^{+0.011}$ & $0.75_{-0.05}^{+0.05}$ & 49.2/27 \\
 7 & $0.81_{-0.11}^{+0.11}$ & $1.75_{-0.08}^{+0.07}$ & $0.29_{-0.02}^{+0.03}$ & $0.168_{-0.009}^{+0.009}$ & $1.13_{-0.05}^{+0.05}$ & 60.5/47 \\
 8 & $0.48_{-0.10}^{+0.10}$ & $1.71_{-0.09}^{+0.10}$ & $0.208_{-0.019}^{+0.021}$ & $0.130_{-0.007}^{+0.007}$ & $0.83_{-0.04}^{+0.04}$ & 85.5/51 \\
\cutinhead{X-44}
 1 & $1.18_{-0.20}^{+0.18}$ & $2.03_{-0.04}^{+0.07}$ & $0.39_{-0.03}^{+0.04}$ & $0.164_{-0.009}^{+0.009}$ & $1.26_{-0.06}^{+0.07}$ & 35.7/34 \\
 2 & $0.83_{-0.11}^{+0.10}$ & $1.74_{-0.04}^{+0.04}$ & $0.330_{-0.019}^{+0.020}$ & $0.190_{-0.007}^{+0.007}$ & $1.28_{-0.04}^{+0.04}$ & 98.6/87 \\
 3 & $0.5_{-0.2}^{+0.2}$ & $1.63_{-0.09}^{+0.09}$ & $0.35_{-0.03}^{+0.05}$ & $0.232_{-0.019}^{+0.020}$ & $1.48_{-0.09}^{+0.09}$ & 20.7/22 \\
 4 & $0.86_{-0.22}^{+0.15}$ & $1.81_{-0.06}^{+0.06}$ & $0.36_{-0.03}^{+0.03}$ & $0.198_{-0.012}^{+0.013}$ & $1.35_{-0.06}^{+0.06}$ & 37.6/42 \\
 5 & $1.1_{-0.6}^{+0.3}$ & $2.2_{-0.3}^{+0.3}$ & $0.30_{-0.07}^{+0.06}$ & $0.111_{-0.017}^{+0.023}$ & $0.87_{-0.07}^{+0.14}$ & 14.8/15 \\
 6 & $0.75_{-0.21}^{+0.18}$ & $1.68_{-0.07}^{+0.09}$ & $0.34_{-0.03}^{+0.04}$ & $0.214_{-0.013}^{+0.013}$ & $1.40_{-0.07}^{+0.07}$ & 41.4/34 \\
 7 & $0.64_{-0.20}^{+0.18}$ & $1.90_{-0.07}^{+0.08}$ & $0.22_{-0.02}^{+0.03}$ & $0.112_{-0.007}^{+0.008}$ & $0.76_{-0.04}^{+0.04}$ & 21.7/29 \\
 8 & $0.86_{-0.15}^{+0.14}$ & $1.96_{-0.07}^{+0.06}$ & $0.30_{-0.02}^{+0.03}$ & $0.141_{-0.008}^{+0.008}$ & $1.01_{-0.04}^{+0.04}$ & 37.1/46 \\
\enddata

\tablecomments{
Col.~(1): Observation index corresponding to Column (1) of Table~\ref{tab:antobs}.
Col.~(2): Absorption column density in units of $10^{21}$~cm$^{-2}$.
Col.~(3): Photon index of the {\tt powerlaw} model.
Col.~(4): Normalization of the {\tt powerlaw} model at 1~keV in units of $10^{-4}$~photons~cm$^{-2}$~s$^{-1}$.
Col.~(5): Absorbed flux in 0.3--10 keV in units of $10^{-12}$~\ergcms.
Col.~(6): Unabsorbed luminosity in 0.3--10 keV in units of $10^{40}$~\ergs.
Col.~(7): Best-fit $\chi^2$ and degrees of freedom.
All errors are at 1 $\sigma$ level. 
}
\end{deluxetable*}

The Antennae (NGC 4038/4039) are a pair of merging galaxies, which have a high star formation rate and harbor a wealth of bright X-ray sources as well as ULXs \citep{zez02a}. With archival \cha\ observations, four ULXs, X-11, X-16, X-42, and X-44, are found to have particularly high luminosity and variability \citep{fen06b}. We proposed six \xmm\ observations of the Antennae (PI: H.\ Feng) in order to investigate spectral variabilities of these ULXs. Another two archival observations are also included \citep{mil04b}.
 
Due to the crowded field and strong background emission from the interstellar medium (ISM), we extracted spectra from smaller regions than usual, which are 12\arcsec\ radius for X-11 and X-16 and 10\arcsec\ radius for X-42 and X-44. As mentioned above, the small extraction region does not affect the inferred flux. The PN data in observations 1, 2, 5, 7, and 8 are not used for X-11, because it lies on the CCD gap. No background subtractions because the small extraction region results in negligible instrumental background.  The ISM does contribute a background, but this varies with position in the galaxy. To account for the ISM emission, we model its spectrum using the thermal plasma model {\tt apec} with parameters adjusted for each source. The distance to the Antennae is controversial: $\sim$20~Mpc from recession velocity, 13.3~Mpc derived by measuring the tip of the red giant branch \citep[TRGB;][]{sav08}, and 22~Mpc based on a type Ia supernova and re-analysis of the TGRB \citep{sch08}. In our previous paper \citep{fen06b}, we adopted a distance of 13.8~Mpc, which is estimated from TRGB \citep{sav04}. Here, we adopt the most recent result of 22~Mpc, which leads to higher luminosities by a factor of 2.5 than before. The absorption column density in our Galaxy is $0.314 \times 10^{21}$~cm$^{-2}$.

The ISM emission for each source was determined iteratively. First, we fit every spectrum using a power-law plus {\tt apec} model with all parameters free except the {\tt apec} abundance (set to solar) and redshift (set to zero). Then, we fixed the plasma temperature for each source at its mean value averaged from different observations and fit the spectra again to find new normalizations. Finally, we fixed the {\tt apec} plasma temperature to the mean temperature and the normalization to the mean normalization.  The resulting {\tt apec} model for each source served as a background spectrum for that source. We note that the final mean values are consistent with the best-fit values at the first step for each single observation, which is reasonable as a result of steady ISM emission. The temperature is 0.43, 0.38, 0.44, and 0.45~keV, and the normalization is 9.6, 5.6, 13, and 3.2 (in units of $10^{-6}$ times the {\tt apec} normalization; see XSPEC manual), respectively for X-11, X-16, X-42, and X-44. According to the quality of the data and small contribution of the ISM emission, slight changes of those values will not affect the best-fit values of other parameters. 

\begin{figure*}
\centering
\includegraphics[width=0.49\textwidth]{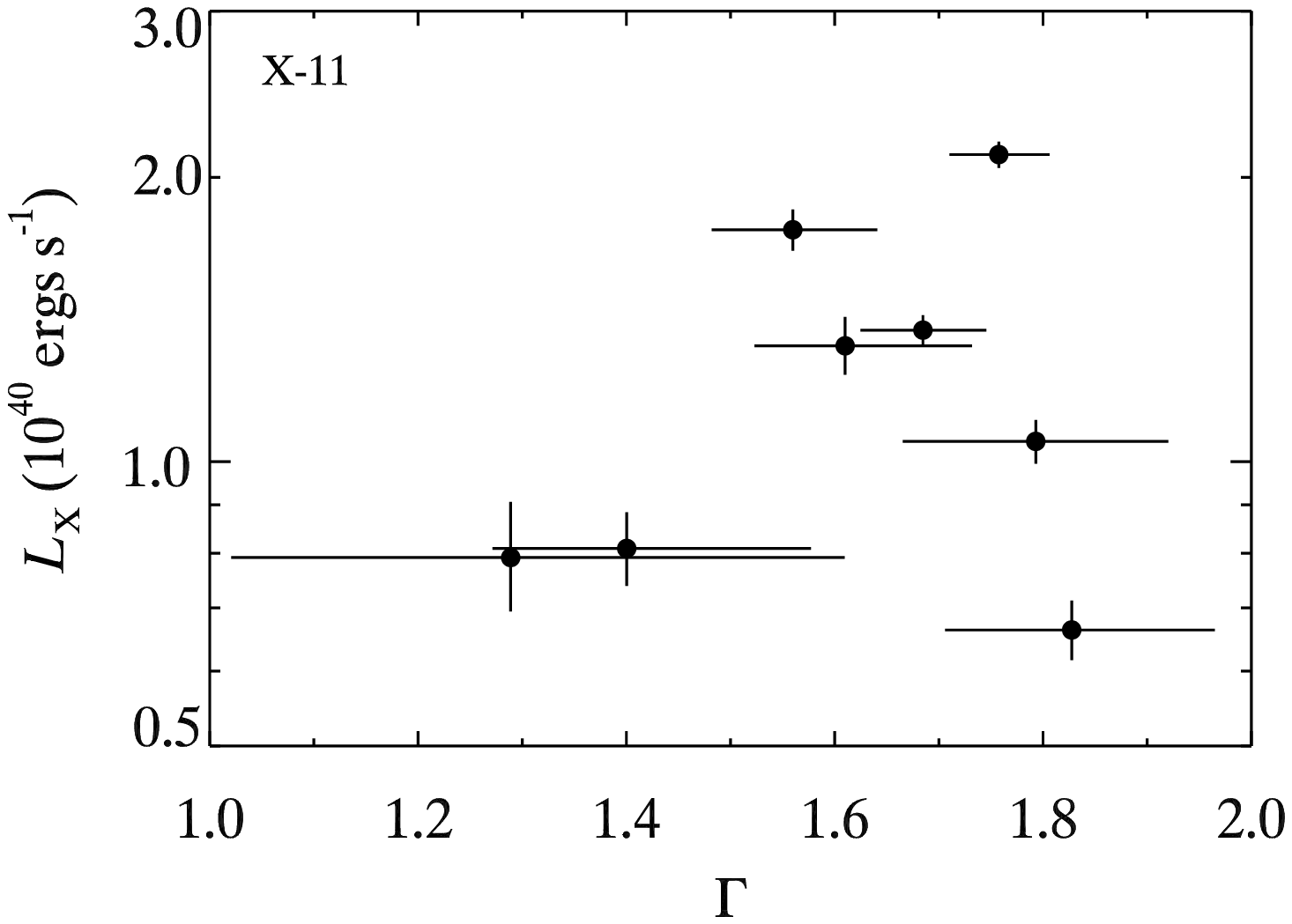}
\includegraphics[width=0.49\textwidth]{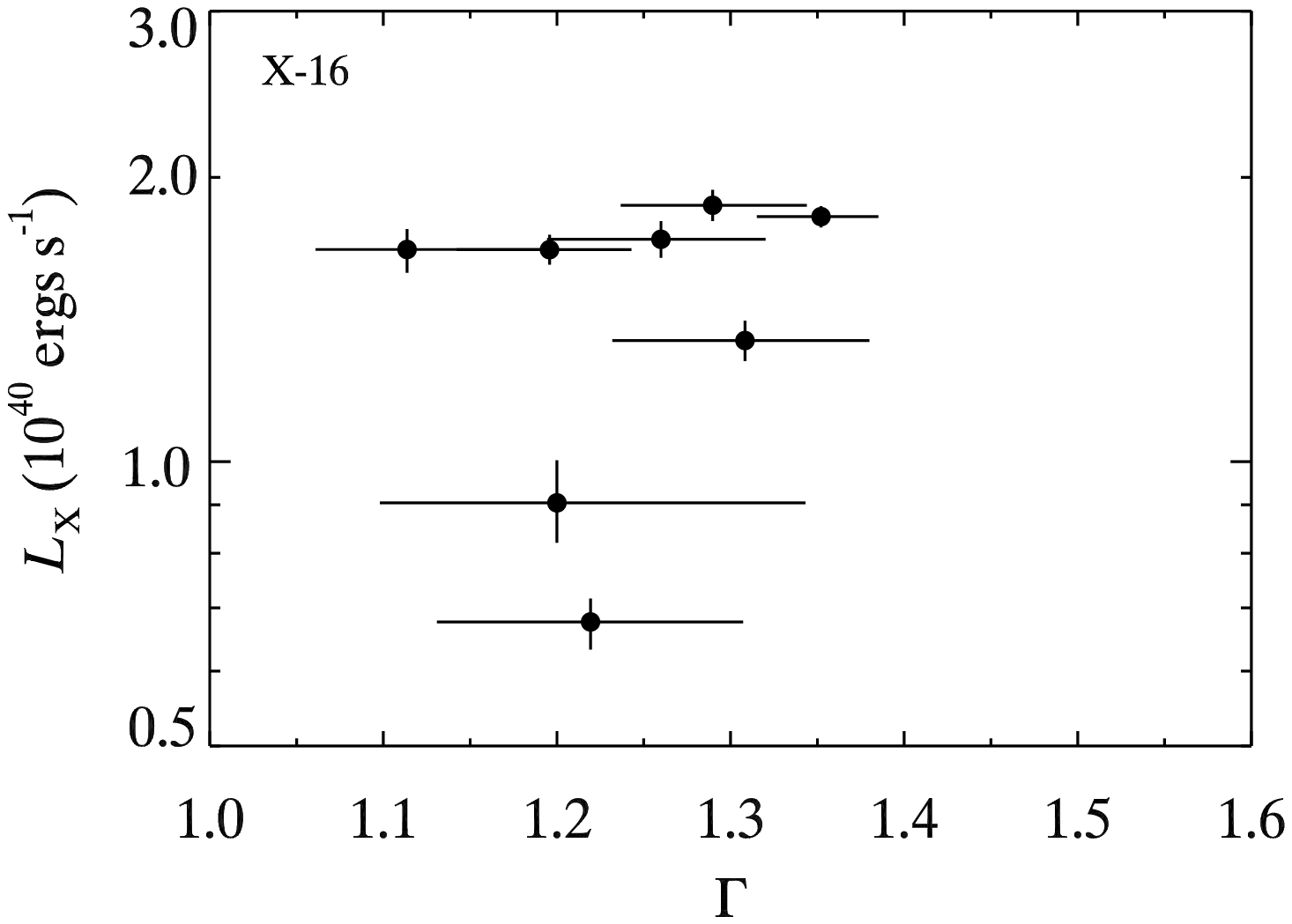}\\
\includegraphics[width=0.49\textwidth]{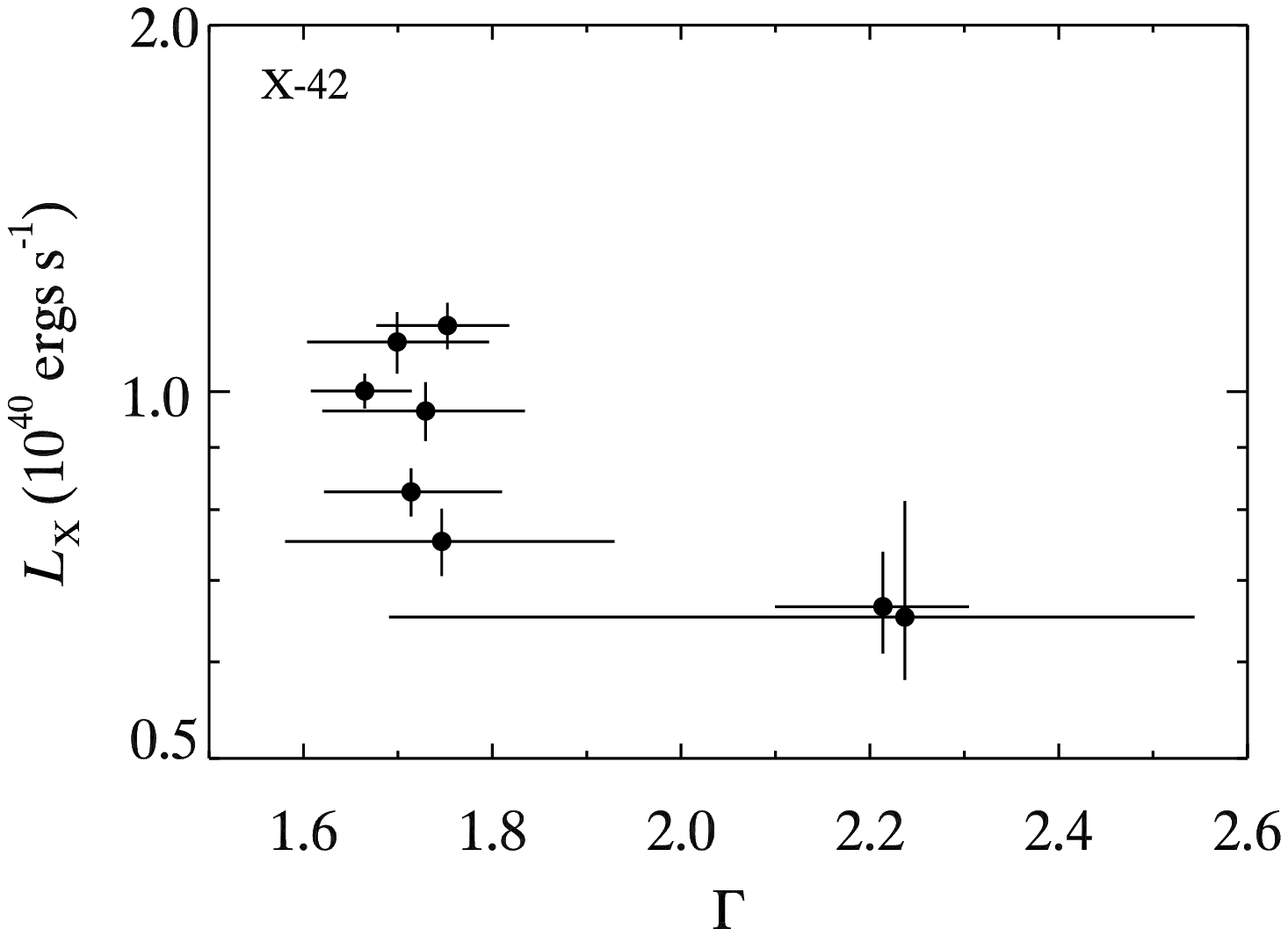}
\includegraphics[width=0.49\textwidth]{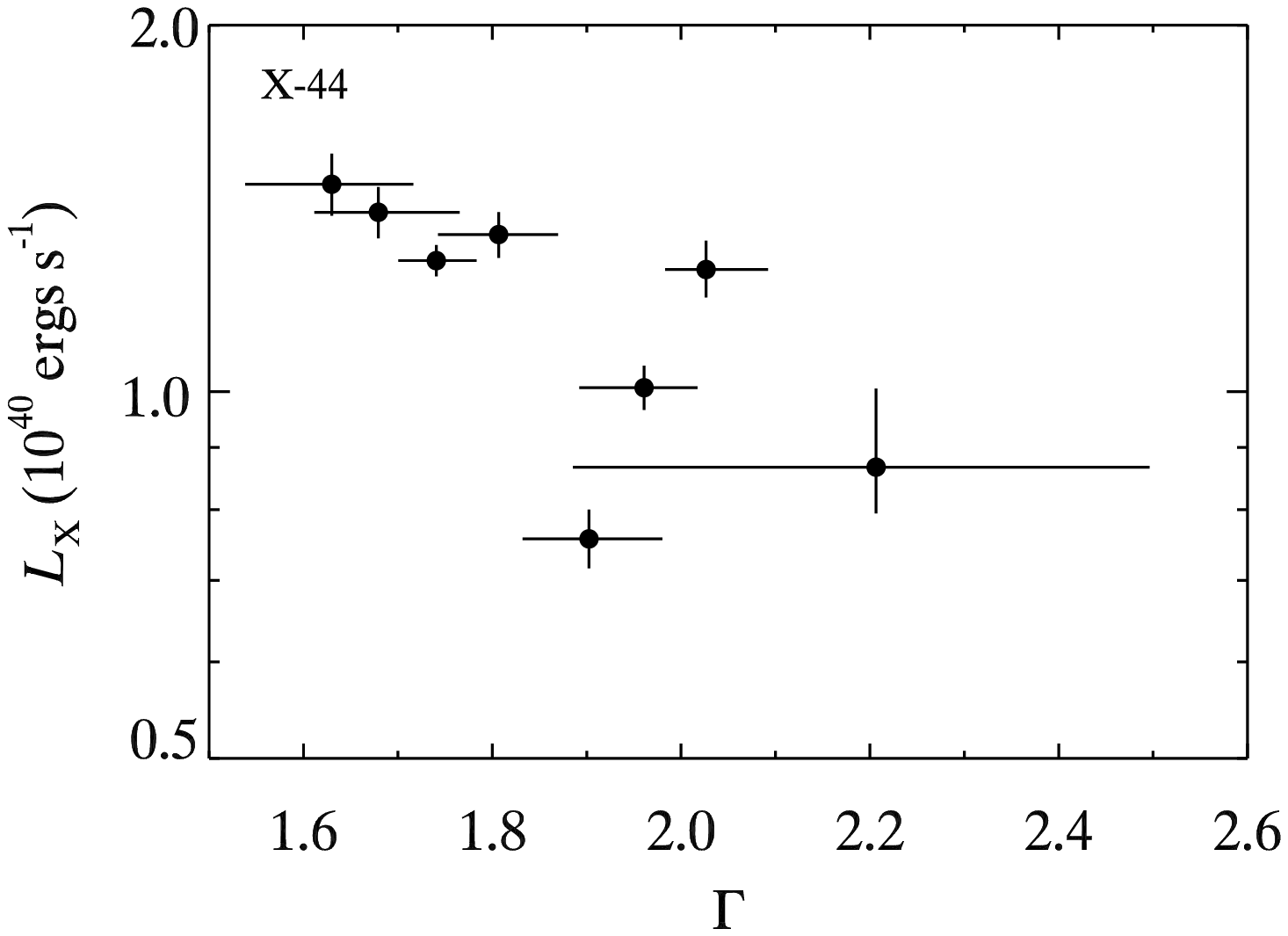}
\caption{
$L_{\rm X}$ versus $\Gamma$ of Antennae X-11, X-16, X-42, and X-44 derived from an absorbed power-law model (data points adopted from Table~\ref{tab:antfit}). 
\label{fig:ant}}
\end{figure*}

We tried a single power-law model subject to interstellar absorption to fit the data. All best-fit parameters are listed in Table~\ref{tab:antfit}. We plotted $L_{\rm X}$ versus power-law photon index $\Gamma$ in Figure~\ref{fig:ant} for the four sources. It is interestingly shown that all sources, especially X-11, X-16, and X-42, display a constant $\Gamma$ despite dramatic changes of the luminosity. X-44 shows some trends of hardening when the source becomes bright, which is consistent with results from previous \cha\ observations \citep{fab03,fen06b}. X-16 exhibits an extremely hard spectrum with $\Gamma=1.1 \sim 1.4$, and a low absorption column density close to the Galactic value.

\section{Discussion}
\label{sec:diss}

In the past decade with {\it Rossi X-ray Timing Explorer} observations, an observational description of spectral evolution in black hole binaries has been established with the identification of four states (quiescent, hard, thermal dominant, and steep power-law) and transitions between them \citep[for reviews see][]{mcc06,rem06}. Each state is characterized by its unique spectral, timing, and multiwavelength properties, and linked to a specific physical configuration of the accretion flow. A standard, phenomenological spectral model consisting of a disk blackbody plus power-law is widely adopted to describe the X-ray spectrum of black hole binaries. Here, we discuss our results on ULXs by comparing them with Galactic black hole binaries in terms of these spectral states.

\subsection{The Cool Disk Model}

The correlation between the luminosity and disk temperature of NGC 5204 X-1 suggests that the soft excess in the spectrum can be interpreted as emission from an accretion disk. The $L \propto T^{2}$ relation found between the luminosity and inner disk temperature is similar to that seen at the highest luminosities observed from XTE~J1550$-$564 in its 1998-1999 outburst \citep{kd04} and from H1743$-$322 in its 2003 outburst \citep{mcc07}.  The effective temperature observed from an accretion disk is modified by `spectral hardening' due to the disk atmosphere.  Using state-of-the-art disk atmosphere models, \citet{mcc08} analyzed the data for XTE~J1550$-$564 and H1743$-$322 and showed that the deviations from the $L \propto T^{4}$ seen at high luminosity when the luminosity is plotted against the observed disk temperature are largely removed when the spectral hardening is taken into account and the luminosity is plotted against the effective disk temperature.  Our analysis presented above, shows that application of a spectral hardening correction to the data for NGC 5204 X-1 produces a similar effect: the luminosity versus temperature relation for the effective disk temperature is close to the $L \propto T^{4}$ relation expected for an accretion disk.  Therefore, we interpret the source emission of NGC 5204 X-1 as due to a standard or quasi-standard accretion disk with the observed spectrum modified by spectral hardening.

The mass of the compact object can be derived under this scenario. Assuming the source peak luminosity, $\sim 10^{40}$~\ergs, is under the Eddington limit, which is $1.3 \times 10^{38}$~\ergs\ for a solar mass black hole, a lower limit on the mass can be calculated as $M > 80$~$M_\sun$. The inner radius derived from the disk model has a maximum of about 1000~km and slowly declines at high temperatures to about 600~km. Similar to XTE~J1550$-$564 \citep{km04}, the maximum $R_{\rm in}$ in such a phase corresponds to the innermost circular stable orbit (ISCO) of the accretion disk, which is 9~km for a  non-spinning black hole of one solar mass. This places a lower limit on the compact object mass of $M > 110$~$M_\sun$. Adopting Equation (9) in \citet{mak00}, substituting $\xi=0.41$ and $T_{\rm eff} = T_{\rm in}/\kappa$ (where $\xi$ is the radius correction for the maximum temperature, and $\kappa$ is the hardening correction which is equal to $f$ in our paper), we have $L = 6.0 \times 10^{37} (M/M_\sun)^2 (T_{\rm eff}/{\rm keV})^4$~\ergs\ for a non-spinning black hole. Fitting to the data of NGC 5204 X-1 gives a best-fit relation $L_{\rm X} = (7.9 \pm 0.8) \times 10^{42} (T_{\rm eff}/{\rm keV})^4$~\ergs, and, thus, a black hole mass of $360 \pm 120$~$M_\sun$. The luminosity, disk size and temperature all point to a consistent result that NGC 5204 X-1 harbors an IMBH with a few hundred solar masses. We note that the derived mass will be larger for a Kerr black hole, which has a smaller ISCO and hotter disk than a non-spinning black hole. We also note that there is a suggestion of high energy spectral curvature in the spectrum indicative of an optically thick corona, which could hide the innermost accretion disk leading to an overestimate of the black hole mass. However, the evidence for curvature is weak in the \xmm\ band, and is only seen in a few bins at energies close to 10~keV. Future observations in the hard X-ray band could address this question.

In Holmberg II X-1, the four data points are not inconsistent with $L \propto T^4$ relation. However, more observations are required to accurately measure the spectral evolution and test the $L \propto T^4$ relation before concrete conclusions can be made.

Interstellar absorption and the cool disk model are competing in the same energy range. As demonstrated in Figure~\ref{fig:ngc5204var}, lower abundances will not change the correlation found for the disk model, but will mildly flatten the luminosity and temperature evolution pattern. Higher abundances will affect in the opposite way. Therefore, for systems like NGC 1313 X-2 where an anti-correlation was found between the luminosity and temperature of the cool disk, changes of the abundance will not help recover the $L \propto T^4$ relation. For systems like NGC 5204 X-1 in which the correlation is found, the abundance is important in determining the power-law slope between $L$ and $T$. For NGC 5204 X-1, near-solar abundance was determined using \xmm\ observations \citep{win07}. For Holmberg II X-1, different abundances are measured using \xmm\ RGS \citep[0.6 solar;][]{goa06} and EPIC \citep[1 solar;][]{win07} data. If the 0.6 solar abundance is true, the slope between $L$ and $T$ will be smaller, which makes it more similar to NGC 5204 X-1.

The cool disk interpretation of the spectrum of IC 342 X-1 seems to be incorrect because the luminosity versus temperature data are strongly inconsistent with the $L \propto T^4$ relation (Figure~\ref{fig:ic342x1}). For the two observations with low luminosities, the confidence levels found by F-test of the need for a cool disk component in addition to the power-law component are 0.77 and 0.9992, respectively for observation 1 and 3. Therefore, the cool disk component in the 3rd observation is significant at 3.3$\sigma$, which firms the anti-correlation between the luminosity and temperature, but more observations are needed to test the results. This low/hot to high/cool evolution derived from the cool disk model is quite similar to the behavior of NGC 1313 X-2 \citep{fen07a}. The best-fit power-law exponent between $L_{\rm X}$ and $T_{\rm in}$ is also similar for the two systems: $-2.8 \pm 0.7$ for IC 342 X-1 and $-3.7 \pm 0.7$ for NGC 1313 X-2. \citet{pou07} predicted an anti-correlation between $L_{\rm X}$ and $T_{\rm in}$ in a model that assumes ULXs are super-Eddington sources similar to SS 433 and the soft excess is due to thermal emission at the spherization radius where outflows start. Both IC 342 X-1 and NGC 1313 X-2 are surrounded by supernova remnant like nebulae \citep{pak02,pak03}, which may be produced by strong outflows from the central source, like W 50 around SS 433. Therefore, IC 342 X-1 and NGC 1313 X-2 may represent a class of sources with super-Eddington accretion.

\subsection{The Power-law/Comptonization Component}

A power-law model usually provides good fits to the high energy part of \xmm\ spectra of ULX, leading to interpretation of the emission as from an optically thin corona. However, in some cases, spectral curvature is evident and a Comptonization model, or equivalently a cutoff power-law model, is favored \citep{fen05,sto06}. Fitting with the Comptonization model requires an optically thick corona in some or all observations of NGC 5204 X-1, Holmberg II X-1, and IC 342 X-1.  The detected spectral curvature starts at energies close to the upper bound of the \xmm\ effective energy range. Therefore, non-detection of an optically thick corona could be a result of the limited energy range, and even with a detection, the significance is usually low due to lack of information above 10~keV. The dominant luminosity of the power-law or Comptonization component in the \xmm\ band suggests that coronal emission is dominant in ULXs. The coronal temperature in some ULXs is much lower than in Galactic black hole binaries, which could be caused by a cooler and more luminous accretion disk in ULXs that seeds and cools the corona \citep{don06,sto06}.

\subsection{The Thermal Dominant State}

In Galactic black hole candidates, the X-ray emission in the thermal dominant state is dominated by a geometrically thin optically thick accretion disk, which can be well described by the standard accretion disk model \citep{sha73}. The $L \propto T^4$ relation in this state is well defined by observations \citep[cf.][]{gie04} and is, perhaps, the best understood emission state in black hole binaries. However, the thermal dominant state rarely appears in ULXs. \citet{kub01} found that the \asca\ spectrum of IC 342 X-2 in the high state could be adequately fitted by a standard accretion disk model, implying that the source was in the thermal dominant state. This is confirmed by the \xmm\ data in a similar energy range of 1--10 keV. However, as shown in Figure~\ref{fig:ic342x2}, no correlation is found between the luminosity and temperature. The four data points appear at two luminosity levels varying by a factor of 4. Fitting with an $L \propto T^4$ relation results in $\chi^2=120.9$ with 3 degrees of freedom, strongly ruling out the model.

\subsection{The Hard State}

The four ULXs in the Antennae and IC 342 X-1 displayed a hard power-law photon index in all or some of the \xmm\ observations. Antennae X-16 may be the most interesting source among them due to its extremely hard spectrum with $\Gamma = 1.1-1.4$, which are consistent with previous \cha\ results \citep{fen06b}. Some Galactic objects with such a hard spectrum are usually highly absorbed hard X-ray objects with wind accretion from a supergiant companion. However, X-16 has a small absorption column density beyond the Milky Way; $N_{\rm H}$ is pegged to the Galactic value in four out of the eight observations. The source flux is highly variable, while its spectrum is constantly hard and the absorption is low. Lack of radio emission rules out that the source emission is due to relativistic beaming along the line of sight such as blazars \citep{zez02b}. An analogue to X-16 is CXO~J024238.9$-$000055 in NGC 1068, which also shows an extremely hard spectrum with little absorption \citep{smi03}. 

Antennae X-11 and X-42 have a constant $\Gamma$ around 1.6--1.8, except for one data point for X-42 in observation 4 with $\Gamma=2.21_{-0.11}^{+0.09}$ that is softer than most others. Antennae X-44 displays a little trend of softening at low luminosities until $\Gamma = 2.0$. IC 342 X-1 shows a flux change by a factor of a few with $\Gamma < 2.1$. These behaviors are consistent with the classification as in the hard state with $\Gamma = 1.5-2.1$ in black hole binaries \citep{mcc06}. The two brightest (in observed flux) ULXs in nearby galaxies are X41.4$+$60 and X42.3$+$59 in the starburst galaxy M82 \citep{fen07b,kaa08}, and also show hard power-law spectra. X41.4$+$60 remains in the hard state even at the highest fluxes observed \citep{kaa08}. The Antennae galaxies and M82 both have high star formation rates, and the ULXs in them share similar properties. \citet{ber08} found ULXs tends to be hard when the luminosity increases. The hard state represents a luminosity of a few or as high as 30 percent of the Eddington limit \citep[e.g., GX 339$-$4;][]{miy08}. Therefore, the mass of the compact object in these ULXs, if they are in the hard state, is at least a few hundred solar masses.

\subsection{The $L - \Gamma$ Correlated Phase And Steep Power-law State}

NGC 5204 X-1 was probably in an intermediate state with hybrid properties from the thermal dominant and steep power-law states during those \xmm\ observations. In a 50~ks \cha\ observation made in 2003, the source made a transition to the steep power-law state with $\Gamma = 2.6$ or 2.9 (depending on models) at a luminosity of a few $10^{39}$~\ergs\ \citep{rob06}. This certainly broke the tight correlation found between $L_{\rm X}$ and $\Gamma$ as shown in Figure~\ref{fig:ngc5204} and is very similar to the behavior of NGC 1313 X-1 \citep{fen06a}. NGC 1313 X-1 also showed a tight correlation between $L_{\rm X}$ and $\Gamma$ at $\Gamma \le 2.4$, and then jumped into the steep power-law state with $\Gamma = 3.1$. Holmberg II X-1 might also have undergone a similar transition from an $L_{\rm X} - \Gamma$ correlated phase to a steep power-law state (see Figure~\ref{fig:hoii}) based on the four observations. Another similarity between Holmberg II X-1 and NGC 5204 X-1 is that they both show correlation between $L_{\rm X}$ and $T_{\rm in}$. This analogy suggests that soft excesses in NGC 1313 X-1 may be due to disk emission around an IMBH and may follow an $L \propto T^4$ evolution.

This $L - \Gamma$ correlated phase could be common in ULXs with an IMBH accretor. It shows hybrid properties from the thermal dominant state (with an $L \propto T^4$ relation), and the steep power-law state (with strong Comptonization). Sources in this phase show strong variability and occasionally make a transition to the steep power-law state. 

\subsection{IC 342 X-2 And The Peculiar Soft Feature}

The source shows a peculiar feature at energies below 1~keV. A collisionally-ionized diffuse gas model ({\tt apec}) could adequately fit this feature. However, we argue that the best-fit parameters derived from the {\tt apec} model are unphysical. The normalization of the {\tt apec} model is $N_{\rm A} = 10^{-14} n_{\rm e}^2 V /(4 \pi D^2)$, where $n_{\rm e}$ is the electron density in cm$^{-3}$, $V$ is the volume in cm$^3$, and $D$ is the distance to the source in cm. The total kinetic energy of the plasma is $E = 3 kT n_{\rm e} V = 6 \times 10^6 kTN_{\rm A} D^2 / n_{\rm e}$~\ergs, where $kT$ is the plasma temperature in keV. Substituting $kT$, $N_{\rm A}$, and $D$, we obtain $E \sim 10^{57}/n_{\rm e}$ or $10^{55}/n_{\rm e}$~\ergs\ for the two observations. These energies are too high for an accreting system even with an IMBH accretor in an extremely dense environment. The variation of the {\tt apec} component also excludes the possibility that the soft excesses could be caused by a supernova remnant or other large scale emitter. We tried other models but none was able to fit the soft excesses as well as a thermal plasma model. The nature of soft spectral feature in IC 342 X-2 is puzzling.

\subsection{Timing Properties}

The lightcurve of Holmberg II X-1 showed variability in the energy range of 1--4.5~keV (see Figure~\ref{fig:hoiiv}). Periodograms were calculated and a major peak around 20~ks was remarkable, which corresponds to multiple spikes in the lightcurve at the same timescale. It is unclear whether or not the variation is periodic; more observations are needed to test that. If this is a periodic variation and modulated by the orbital motion of the binary, then the observed timescale could be the orbital period. The density of the companion star can thus be inferred under this scenario assuming Roche-lobe overflow as $\rho = 110 P_{\rm hr}^{-2}=3.6$~(g~cm$^{-3}$), corresponding to a low mass late type main sequence star (e.g., M0~V). Unevolved low mass companions have not yet been discovered in black hole binaries, but exist in ultra-compact binaries with a neutron star accretor. It is inconsistent with the general picture that ULXs in star forming regions are usually high mass X-ray binaries \citep[e.g.,][]{fen06b}. However, ULXs in elliptical galaxies, especially those associated with globular clusters, must be low mass X-ray binaries. On the other hand, if the variability is quasi-periodic oscillations, it has a central frequency much lower than those found in other ULXs \citep{str03,str07} and would be a good argument for an IMBH. Therefore, interesting consequences are expected if the nature of this 20~ks variability are revealed with future observations.

We report no other significant short term timing noise from these observations due to the limited sensitivity, though it is expected in the hard state and sometimes in the steep-power law state. 

\section{Conclusion}
\label{sec:con}

In this paper, we analyzed the spectral evolution of NGC 5204 X-1, Holmberg II X-1, IC 342 X-1 and X-2, the Antennae X-11, X-16, X-42, and X-44. New results obtained in this paper and those reported in the literature (including ULXs in M82 and NGC 1313) were discussed together. The main conclusion of the paper is to report two major spectral classes in ULXs: one is the hard state and the other is the $L - \Gamma$ correlated phase. 

Many ULXs, including sources in M82 and the Antennae, are found in the hard state, in which the source shows strong flux variability but the spectral index remains hard and constant. From analogy with the behavior of Galactic black hole binaries in the hard state, the emission of these ULXs should be sub-Eddington and their masses are estimated to be at least a few hundred solar masses. The Antennae X-16 is an extremely hard ULX, with a power-law photon index $\Gamma = 1.1 - 1.4$; sources with such a hard spectrum, low absorption, and high luminosity are rare and difficult to understand.

The second class of ULXs show behaviors including: (i) the luminosity is positively correlated with the power-law photon index, (ii) the source sometimes breaks the correlation and jumps into the steep power-law state, and (iii) the luminosity seems to vary with the 4th order of the disk temperature indicating that the emission is from an accretion disk. Sources in this class includes NGC 5204 X-1, Holmberg II X-1, and NGC 1313 X-1. We interpret these behaviors as a hybrid between the thermal-dominant and steep power-law states. Sources in this class are thought to contain compact objects more massive than stellar-mass black holes inferred from their luminosity versus disk temperature pattern.

Some sources, like IC 342 X-1 and NGC 1313 X-2, show a negative correlation between the luminosity and disk temperature. If the negative correlation can be confirmed with future observations and commonly found in other ULXs, they may represent a third class of ULXs with super-Eddington emission.

Also, we report a peculiar soft excess in IC 342 X-2, which is hard to explain, and strong variation on the timescale of 20~ks from Holmberg II X-1. More observations are needed to test these results.
 
\acknowledgments We thank Shane Davis for helping with the hardening correction, and the anonymous referee for helpful comments that have improved the paper. We acknowledge funding support from NASA grants NNX07AQ54G and NNX08AJ26G.

\end{document}